\documentclass[a4paper,noarxiv,onecolumn,accepted=2025-08-22]{quantumarticle}

\usepackage[utf8]{inputenc}
\usepackage[english]{babel}
\usepackage[T1]{fontenc}
\usepackage[ruled,vlined]{algorithm2e}

\usepackage{graphicx}
\usepackage{amsmath}
\usepackage{bbold}
\usepackage{epsfig}
\usepackage{mathtools}

\usepackage{hyperref}
\usepackage{breakurl}
\usepackage{listings}
\usepackage{pgfplots}
\usepackage{soul}
\usepackage[compatibility=false]{caption} 
\newcommand{\pr}[1] {{\color{black}#1 }}
\newcommand{\prr}[1] {{\color{black}#1 }}

\title{Batched Line Search Strategy for Navigating through Barren Plateaus in Quantum Circuit Training}

\author{Jakab Nádori}
\affiliation{Department of Physics of Complex Systems, E\"otv\"os  Lor\'and  University, Pázmány Péter sétány 1/a, Budapest, 1117, Hungary}

\author{Gregory Morse}
\affiliation{Department of Programming Languages and Compilers, E\"otv\"os  Lor\'and  University, Pázmány Péter sétány 1/a, Budapest, 1117, Hungary}

\author{\prr{Barna Fülöp Villám}}
\affiliation{Department of Physics of Complex Systems, E\"otv\"os  Lor\'and  University, Pázmány Péter sétány 1/a, Budapest, 1117, Hungary}

\author{Zita Majnay-Takács}
\affiliation{Department of Programming Languages and Compilers, E\"otv\"os  Lor\'and  University, Pázmány Péter sétány 1/a, Budapest, 1117, Hungary}

\author{Zoltán Zimborás}
\affiliation{QTF Centre of Excellence, Department of Physics, University of Helsinki, Helsinki, Finland}
\affiliation{Wigner Research Center for Physics, P.O. Box 49, 1525 Budapest, Hungary}
\affiliation{Algorithmiq Ltd, Kanavakatu 3C 00160 Helsinki, Finland}

\author{Péter Rakyta}
\affiliation{Department of Physics of Complex Systems, E\"otv\"os  Lor\'and  University, Pázmány Péter sétány 1/a, Budapest, 1117, Hungary}
\affiliation{Wigner Research Center for Physics, P.O. Box 49, 1525 Budapest, Hungary}
\email{peter.rakyta@ttk.elte.hu}

\begin{document}

\maketitle 

\begin{abstract}
Variational quantum algorithms are viewed as promising candidates for demonstrating quantum advantage on near-term devices.
These approaches typically involve the training of parameterized quantum circuits through a classical optimization loop. However, they often encounter challenges attributed to the exponentially diminishing gradient components, known as the barren plateau (BP) problem.
This work introduces a novel optimization approach designed to alleviate the adverse effects of BPs during circuit training. 
In contrast to conventional gradient descent methods with a small learning parameter, our approach relies on making finite hops along the search direction \pr{determined on a randomly chosen subsets of the free parameters}.
The optimization search direction, together with the range of the search, is determined by the distant features of the cost-function landscape.
This enables the optimization path to navigate around barren plateaus without the need for external control mechanisms.
We have successfully applied our optimization strategy to quantum circuits comprising \prr{$21$} qubits and $15000$ entangling gates, demonstrating robust resistance against BPs. Additionally, we have extended our optimization strategy by incorporating an evolutionary selection framework, enhancing its ability to avoid local minima in the landscape.
The modified algorithm has been successfully utilized in quantum gate synthesis applications, showcasing a significantly improved efficiency in generating highly compressed quantum circuits compared to traditional gradient-based optimization approaches.
\end{abstract}

\section{Introduction}

The Variational Quantum Eigensolver (VQE) originally formulated by Peruzzo et al. \cite{Peruzzo2014} and McCLean et al. \cite{McClean_2016}, has garnered considerable attention within the research community over the past two decades \cite{TILLY20221}. 
VQE uses the variational principle to calculate the ground state energy of a Hamiltonian, a fundamental problem in quantum chemistry and condensed matter physics. Conventional computing methods face limitations in accuracy and efficiency, due to the exponentially growing complexity of the quantum state used to describe many-body systems.
VQE, on the other hand, is expected to be efficient to model these complex wave-functions in polynomial time, making it one of the most promising near-term applications for quantum computing. 
In addition, variational quantum algorithms have demonstrated some degree of resilience to noise \prr{in} quantum hardware, making them a good candidate to deliver reasonable calculation results even on near-term quantum computers.
Consequently, VQE stands as a prominent example of noisy algorithms. In its most general form, it seeks to compute an upper bound for the ground-state energy of a Hamiltonian, typically \prr{providing} the first step to determine the energetic properties of molecules and materials. Potential applications of VQE are therefore extensive, encompassing areas such as drug discovery \cite{8585034,Blunt2022}, material science \cite{Lordi2021}, chemical engineering \cite{Kandala2017,doi:10.1126/science.abb9811,Cao2019}, quantum optimization \cite{Harrigan2021,PRXQuantum.1.020304}, and quantum machine learning \cite{Havlicek2019,Johri2021}.

While the training process in variational quantum algorithms is proven to be NP-Hard \cite{PhysRevLett.127.120502}, the literature highlights four primary areas of research aimed at bringing VQE closer to practical real-world applications. These areas encompass:
(i) Development of optimal measurement techniques to minimize the required number of circuit repetitions for VQE execution.
(ii) Establishment of extensive parallelization methods across multiple quantum computers.
(iii) The creation of strategies to address potential challenges associated with vanishing gradients during the optimization process for larger systems.
(iv) The exploration of tailored error mitigation approaches specifically designed for the VQE algorithm.
The solutions to these open research questions will show the pathways for VQEs to attain quantum advantage, particularly as quantum computing hardware scales up and noise levels hopefully diminish.

In this study, we address a common pitfall encountered in variational quantum algorithms during parameter optimization, known as the barren plateau (BP) problem, as described in prior research \cite{PhysRevA.92.042303,McClean2018,Cerezo2021}. 
The BP arises due to the phenomenon where the gradients of the cost function vanish exponentially as the number of qubits in the quantum model increases. 
For cost functions that exhibit BPs, exponentially many measurements may be required to determine the minimization direction in gradient-based optimization.
This challenge in scaling applies not only to gradient-based optimization but also to derivative-free optimization methods, as discussed \prr{by} Arrasmith et al. \cite{Arrasmith2021effectofbarren}, and remains unresolved even with the application of higher-order optimization techniques \cite{Cerezo_2021}.
Research efforts have linked the BP problem to the overall expressiveness of the variational quantum circuit ansatz \cite{Holmes2016}.
The degree of entanglement of the wave-function \cite{PRXQuantum.2.040316,PhysRevResearch.3.033090,PhysRevLett.126.190501,Wang2022}, {the structure of the circuit ansatz \cite{CerveroMartin2023barrenplateausin}}, and the non-locality of the cost function \cite{Cerezo2021,Uvarov_2021,PhysRevLett.128.180505} also have high significance on the likelihood of encountering BPs during the optimization process.
These findings imply that BPs are a fundamental property of variational quantum algorithms at scale, needing extra measures to mitigate their impact.

Numerous methods have been proposed to address this issue, but it remains uncertain whether these approaches can resolve the BP problem.
Early research initiatives focused on strategies to circumvent BPs during the initialization phase of algorithms. For instance, Grant et al. \cite{Grant2019initialization} suggested initializing the circuit with blocks containing identity gates. Skolik et al. \cite{Skolik2021} proposed optimizing the circuit layer by layer, {the work of Liu et. al. used a method based on random gate activation to sequentially increase the expressiveness of the circuit \cite{PhysRevResearch.5.L032040}}, while Dborin et al. \cite{Dborin_2022} recommended employing a matrix product state ansatz for the optimization to reduce entanglement.
An alternative strategy was formulated in the study of \cite{PRXQuantum.3.010313} to take advantage of correlated or restricted single-qubit rotation parameters (or direction).
An alternative approach utilizing the method of natural gradient  \cite{haug2021optimal,PhysRevResearch.2.043246} was also investigated in the literature and has been found to be a promising strategy compared to conventional gradient-based optimizers.
Though, the individual optimization steps turned out to be quite expensive and it's tolerance against BPs was not addressed.

Taking a different approach, the relationship between the occurrence of BPs and the structure of the cost function has been explored in prior publications \cite{Cerezo2021} and \cite{Uvarov_2021}. 
Additionally, the introduction of concepts like entanglement-induced BPs \cite{PRXQuantum.2.040316} and noise-induced BPs \cite{Wang2021} have added new routes to the discussion.
The relation between BPs and entanglement has lead to various proposals that suggest controlling entanglement to mitigate BPs \cite{Kim_2022,PhysRevA.106.052424,PhysRevResearch.3.033090,10.21468/SciPostPhys.14.6.147}. 
Furthermore, the idea of weak BPs, as defined for a subset of qubits \cite{PRXQuantum.3.020365}, has been employed to detect early indicators of BPs and prevent them from occurring. 
In addition, \cite{PRXQuantum.3.020365} has demonstrated that entanglement-induced BPs and BPs arising from local cost functions are essentially the same. Therefore, strategies to avoid entanglement-induced BPs are equivalent to strategies for avoiding BPs associated with local cost functions.
It is important to highlight that all of the suggested methods for mitigating BPs primarily focus on avoiding entering onto a BP in the first place. There is currently no known methodology to follow when the optimization process is already in the midst of a barren plateau. 
One might even argue that barren plateaus can be utilized to characterize the trainability of variational quantum algorithms: an algorithm shows better trainability when experiencing less impact from BPs.

\paragraph*{Contributions:} In this study, we investigate the possibility to engineer a BP resistant optimization technique, showing lower degree of dependence on BP indicators like the vanishing gradient components or high entanglement entropy. 
{In contrast with \cite{PRXQuantum.3.020365} proposing adaptive decrease of the learning rate, we present a counter-intuitive approach implementing finite hops during the optimization. 
Integrating exact or inexact line search \cite{Nocedal2006} is a commonly utilized technique in classical optimizations to improve their efficiency.
In this work we show that line search proves to be exceptionally efficient also in navigating the optimization trajectory \pr{into narrow valleys}.
\pr{Secondly, we show that narrowing down the number of parameters to be updated in each iteration significantly improves the optimization process in the first stage of the execution, especially when initiating the circuit ansatz with identity qubit rotations. 
We attribute this numerical observation to the limited expressiveness of the initial circuit when varying only a small subset of the free parameters in each iteration.}
Additionally, the approach seamlessly integrates into the standard VQE framework, as it only requires the evaluation of the target cost function to be optimized.
We further extend the optimization strategy by an evolutionary selection scheme proven to increase the efficiency in avoiding local minima of the optimization landscape \cite{Failde2023}.
The idea to use evolutionary machine learning practices has been proven in several scenarios, trying to increase the training capabilities of variational quantum algorithms in the context of quantum architecture search \cite{Du2022,10.1145/3588983.3596692,PhysRevLett.116.230504,PhysRevApplied.16.044039,PRXQuantum.2.020310,Cerezo_2021,10.1145/3520304.3534012}
We demonstrate the efficiency of our approach through several applications, including \prr{the approximation of} the ground state of XXX Heisenberg and SYC models, or quantum gate compilation of deep circuits.
Among the key advantages of our algorithm, we can also mention that the computations cost of a single iteration do not exceed the complexity of a conventional gradient based approach optimizing over the same number of parameters.
By evaluating distant features of the cost-function landscape, the optimization strategy can avoid flat areas while not stepping over narrow valleys either.
Our work focusing on VQEs readily extends to other variational hybrid algorithms as well, such as quantum machine learning, quantum optimization or variational time evolution.}
The developed optimization algorithm can be also ported to realistic experiments performed on quantum processors, since it relies on the evaluation of the cost function with simple phase shifts applied on the parameters.
We performed our numerical experiments using quantum computer simulations implemented in the SQUANDER package \cite{SQUANDER_github}.

The structure of the paper is organized as follows. 
In the first section following the introduction we overview the concept of the VQE algorithm.
Subsequently, we present the technical background of our novel optimization method exploiting long-range properties of the optimization landscape to determine the search direction and performing evolutionary selection strategy to further increase the success rate to find the global minimum. 
In the third section we outline the concept of the entanglement entropy -- BP link, using the second-Rényi entropy for BP inference.
In the fourth section we present our numerical results on training quantum circuits to approximate the ground states of the addressed Heisenberg and SYC models and to compile quantum programs.
Finally, we conclude our work in the last section.

\section{Brief description of the VQE algorithm}

As implied by its name, VQE aims to approximate the ground state of a quantum system that is challenging to simulate using classical hardware.
Thus, the first step in the VQE is to define the system for which we want to find the ground state. 
By providing a Hamiltonian $\hat{H}$ and a trial wave-function $|\Psi\rangle$, the ground state energy associated with this Hamiltonian, $E_0$, bounded by 
\begin{equation}
    E_0 \le \frac{\langle\Psi |\hat{H}| \Psi \rangle}{\langle\Psi |\Psi\rangle} \label{eq:VQE1}
\end{equation}
The primary objective of VQE is to determine a parameterization for $|\Psi\rangle$ that minimizes the expectation value of $\hat{H}$. 
This expectation value serves as an upper bound for the ground state energy \cite{arfken1985rayleigh} and, in an ideal scenario, should be practically identical to it, meeting the desired level of precision.
Therefore, VQE operates by variational optimization of the parameters within the quantum circuit ansatz. 
\begin{figure}
     \centering
     \includegraphics[width=0.5\textwidth]{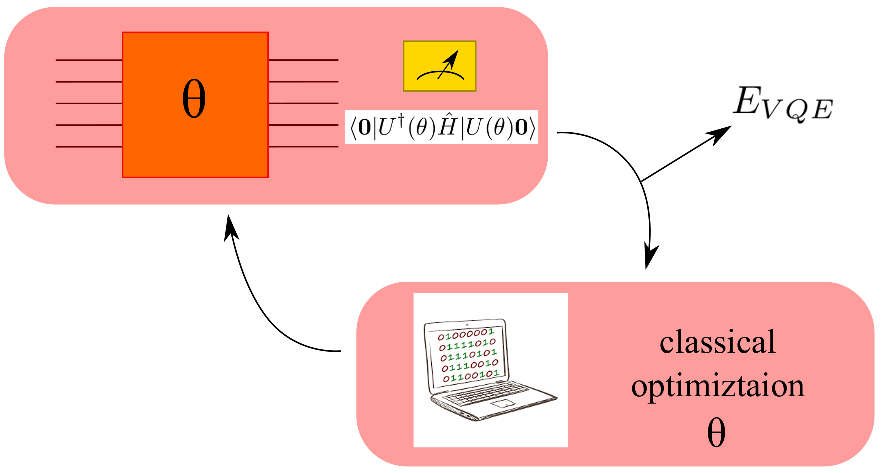} 
     \caption{Schematic process of the VQE algorithm. The expectation \prr{values of} the Hamiltonian is evaluated using the quantum processor parameterized with a parameter vector $\boldsymbol{\theta}$. 
     The VQE energy is optimized via classical computer by updating the parameter vector $\boldsymbol{\theta}$ provided for the quantum circuit in each iteration.}
     \label{fig:vqe}
 \end{figure}

To transform the minimization task into a quantum computer-executable problem, the first step is to define an ansatz quantum circuit used to prepare the trial wave-function on a quantum processor. Therefore, we represent $|\Psi\rangle$ as the result of applying a generic parameterized unitary operation $U(\boldsymbol{\theta})$ to an initial state of $N$ qubits, with the parameter set denoted as $\boldsymbol{\theta}$, taking values \prr{from} the range $\left(0, 2\pi\right]$. For simplicity, the qubit register is typically initialized as $|0\rangle^{\otimes N}$, denoted as $|\mathbf{0}\rangle$.
The VQE procedure is illustrated in Fig.~\ref{fig:vqe}.
Given that $|\Psi\rangle$ must be a normalized wave-function due to the unitary evolution, we can now formulate the VQE optimization problem as follows, omitting the normalization factor from Eq.~(\ref{eq:VQE1}):
\begin{equation}
    E_{VQE} = \min\limits_{\theta} \langle\mathbf{0} | U^{\dagger}(\theta) \hat{H}| U(\theta) \mathbf{0} \rangle  \label{eq:EVQE}
\end{equation}
Equation (\ref{eq:EVQE}) is referred to as the cost function in the context of the VQE optimization problem. 
This terminology is adopted from the fields of machine learning and mathematical optimization literature.

To implement the VQE algorithm on real quantum processors, a typical approach is to expand the Hamiltonian in terms of quantities that can be directly measured on a quantum device. 
This representation is often achieved by expressing the Hamiltonian as a weighted sum of Pauli spin operators.
Observables suited for immediate measurement on a quantum device can be formed by tensor products of spin operators. 
These observables can be defined as Pauli strings denoted as $\hat{P}_{\alpha} \in {I, X, Y, Z}^{\otimes N}$, where $N$ represents the number of qubits utilized in the quantum register.
The Hamiltonian can then be reformulated as follows:
\begin{equation}
    \hat{H} = \sum\limits_{\alpha}^{\mathcal{P}}{w_{\alpha} \hat{P}_{\alpha}}
\end{equation}
with $w_{\alpha}$ being a set of weights, and $\mathcal{P}$ the number of Pauli strings in the Hamiltonian. 
Equation (\ref{eq:EVQE}) then becomes
\begin{equation}
    E_{VQE} = \min\limits_{\theta} \sum\limits_{\alpha}^{\mathcal{P}}{w_{\alpha} \langle\mathbf{0} | U^{\dagger}(\theta) \hat{P}_{\alpha}| U(\theta) \mathbf{0} \rangle }
\end{equation}
where the terms $E_{P_a} = \langle\mathbf{0} | U^{\dagger}(\theta) \hat{P}_{\alpha}| U(\theta) \mathbf{0} \rangle$ \prr{correspond} to the expectation value of a Pauli string $\hat{P}_{\alpha}$ and can be measured by a quantum device, while the summation of these terms \prr{is} calculated using a classical machine.
Generally, to evaluate $E_{VQE}$ with the aid of a quantum chip it is required to sample the expectation values of the Pauli strings several times.
Typically, in order to attain a precision of $\epsilon$ for the expectation value of an operator, it is necessary to conduct approximately $\mathcal{O}(1/\epsilon^2)$ iterations of the circuit execution, each terminated by a measurement at the end \cite{McClean_2016}. 
Therefore, to increase the efficiency of the VQE execution measurement strategies are studied to minimize the number of the required repetitions \cite{PhysRevA.92.042303,Rubin_2018,arrasmith2020operatorsamplingshotfrugaloptimization}.

Once the parameters $\mathbf{\theta}$ have been successfully optimized, the trial state $|\Psi\rangle$ serves as a model for the system's ground state wave-function under investigation. Throughout the optimization process, the ansatz parameters must undergo iterative updates until convergence is achieved. Therefore, the selection of a classical optimization algorithm plays a crucial role. Firstly, it directly influences the number of measurements needed for each optimization step and the total number of iterations required to reach convergence. Secondly, specific optimization strategies have the potential to partially mitigate particular optimization challenges, such as the BP problem.
Though, a full-fledged solution to handle BPs during the optimization process still forms an open challenge for researchers.

\section{Evolutionary optimization technique to train quantum circuits}

Evolutionary algorithms \cite{265956,10.1162/evco_a_00325} apply the principles of evolution found in nature to find the most optimal configuration of properties to accomplish specific tasks.
They are often applied in scenarios where solving problem in polynomial time is not feasible, like in the case of combinatorial optimization problems, leading to NP-hard tasks at scale. 
In this sense, evolutionary algorithms optimize among different possibilities in the space of solution candidates, which are encoded in specific properties of the individuals, referred to as \emph{agents} in the \prr{following}.  
 Evolution is therefore based on recombination over a group of agents, accompanied  with random mutations. 
 Natural selection between \prr{st{the}} agents is performed according to the optimization criteria, encoded in the cost function. 
This process is repeated until the individuals can not improve any more in solving the optimization problem.

The agents explore the parameter space to find a favorable route that eventually decreases the cost function the most. 
A possible and numerically efficient scenario might be to follow a trajectory traced by the optimization of the cost function with respect to a small subset of the free parameters.
In other words, the agents perform in each iteration an exact line search 
along the direction determined in a randomly selected parameter-subspace while keeping the rest of the parameters constant.
Studies of Refs. \cite{PhysRevResearch.2.043158,PhysRevResearch.3.033083} and \cite{Ostaszewski2021structure} showed that by meeting certain conditions in the parameterization of the quantum gates, the single parameter dependence of the energy landscape in a VQA algorithm can be described by a simple sinusoidal function with a period of $2\pi$. 
This finding has inspired the creation of two high performance optimization \prr{algorithms} demonstrated in this work.
First of all we should mention that merely three calls to the cost function are sufficient to find the exact minimum with respect to a single parameter. 
Secondly, global-scaled properties on the optimization landscape can be collected by sampling the cost function along single-parameter directions, leading to the formulation of a novel strategy on deciding the search direction -- and its range -- during the optimization iterations.
These mathematical findings enable a monotone step-by-step reduction of the cost function by iteratively selecting a new subset of parameters during the optimization process.
The result of the procedure is determined by both the starting point in the parameter space and the sequence of the chosen parameters.
Unfortunately, neither the optimal starting point nor the most optimal parameter subset sequence is known in advance.
While a predefined sequence of parameters, such as an iterative parameter sweep following a periodic pattern, might prematurely lead to convergence in most cases, a well-chosen sequence of optimized parameters could significantly enhance the overall optimization efficiency.
In this work we show via numerical experiments, that \prr{probability-based} approaches promise a high success rate in finding a sub-optimal sequence of the optimized parameters.

\paragraph*{Evolutionary selections:}

The algorithm described in this study is based on a highly-efficient parameter-wise optimization routine and on an evolutionary selection of the sequence of the parameters over which the cost function is minimized.
In our algorithm, each agent randomly selects the next parameters to be optimized by sampling from a uniform distribution. 
After optimizing the cost function with respect to the chosen parameters, agents pick up a next subset to optimize, again via uniform distribution sampling.

This repetitive process charts a path in the parameter space, over which the cost function is being optimized.
Upon completing a specified number of iterations (set to a value between $200$ and $20000$ in our reference implementation within the SQUANDER package), the evolution of the agents are compared to each other and the most successful agent (returning the most favorable cost function) is selected. 
We refer to this step in the algorithm as \emph{agent synchronization protocol}.

At this stage, the individual agents are randomly assigned to either continue their optimization path with a probability of $p_{exploration}$ or start a new route from the actual position of the most successful agent. The latter option has a probability of $1-p_{exploration}$. In the reference implementation of the algorithm $p_{exploration}=0.2$ was chosen as default.
\begin{figure}
     \centering
     \includegraphics[width=0.95\textwidth]{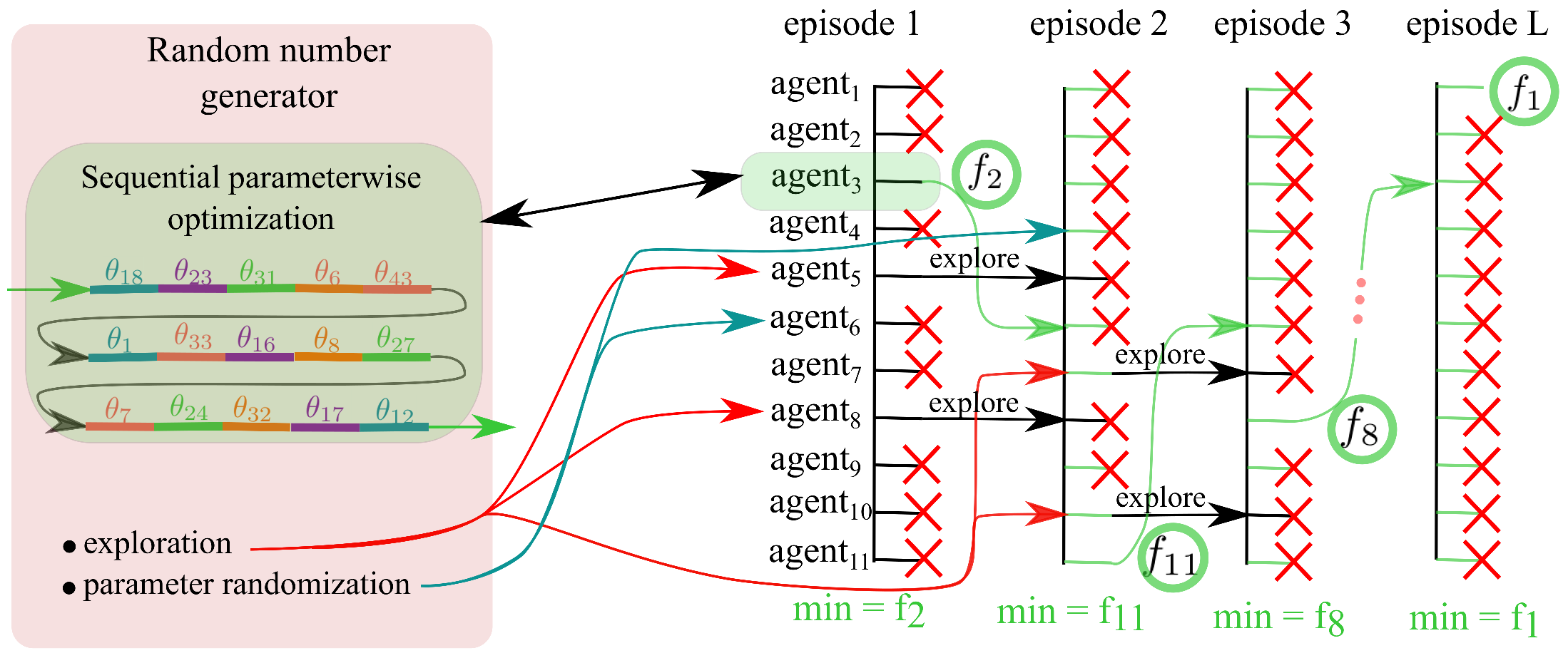} 
     \caption{Illustration of the evolutionary algorithm in a corner case of a single-parameter optimization per iteration step. 
     In each episode the agents perform parameter-wise optimization along a path defined by a randomly chosen sequence of parameters.
     At the end of an epoch the most successful agent is selected and its state is spawned over the remaining agents for the next episode.
     The remaining agents are either terminated (indicated with red crosses), or randomly selected to keep their state and resume their evolution to further explore the parameter space (see the red arrows indicating the chosen agents).
     Some agents are randomly chosen to implement a randomization on their state (see the blue arrows indicating the chosen agents). 
     The figure illustrate a specific example with $11$ agents, resulting in a sequence of best solutions $f_2 > f_{11} > f_8 > \dots > f_1$, with the final result $f_1$ after $L$ episodes.}
     \label{fig:evol_alg}
 \end{figure}
 
We split the optimization algorithm into episodes: each episode terminates with the described agent synchronization protocol and encompasses the evolution of the agents between two synchronization points. To further enhance the success rate \prr{by} avoiding local minima of the energy landscape, the parameter vector associated with a given agent undergoes a perturbation with a probability $p_{randomization}$ right after the agent-synchronization protocol.
This action causes an adjustment in the elements of the parameter vector by a randomly selected value from the range $[-r, r]$. In our reference implementation of the algorithm $p_{randomization}=0.2$, and $r=0.3$  was chosen as default.
The described algorithm is schematically visualized in Fig.~\ref{fig:evol_alg}.

\paragraph*{Single parameter optimization:}
 
The algorithm's performance is highly influenced by the efficiency of the line search that is commonly enhanced by performing in an inexact manner\cite{Nocedal2006}.
Instead of finding the exact minimum in the search directions, an inexact line search terminates after finding an acceptably good point providing sufficient decrease in the cost function.
In our case, however, the sinusoidal structure of the landscape enables to design an efficient method to find the exact minimum with respect to the chosen parameter just in few evaluation of the cost function.
Firstly, the circuit can be decomposed into a sequence of single-parameter unitaries, each having a form $U_i=\textrm{exp}(\theta_i \hat{H}_i)$, where $\hat{H}_i$ is a hermitian matrix generating a rotation. 
Scaling this matrix allows each unitary $U_i$ to be adjusted to have a period of $2\pi$.
Secondly, if $\hat{H}_i$s are full-rank matrices, the unitaries $U_i$s \prr{cannot} facilitate controlled rotations.
Consequently, parameter-free entangling gates are added to the circuit which do not affect the sinusoidal nature of the single-parameter dependence of the cost function which reads as
\begin{equation}
   E_{VQE} = \kappa\cdot\sin(2p_i+\xi) + C \;. \label{eq:3point}
\end{equation}
The derivation of this equation is straightforward as the state generated by the circuit can be given in the form
\begin{equation}
    |\Psi\rangle  = \cos(p_i)|a\rangle + \sin(p_i)|b\rangle
\end{equation}
with respect to the parameter $p_i$, where $|a\rangle$ and $|b\rangle$ are normalized quantum states. 
By evaluating the expectation value $\langle\Psi |\hat{H}| \Psi \rangle$ we end up with Eq.~(\ref{eq:3point}) by introducing suitable constants $\kappa$, $\xi$ and $C$.
In this work we generalize this equation to cases when parametric controlled rotation gates are also incorporated in the circuit, particularly significant in certain approaches of quantum gate compilations \cite{squander2, Nemkov2023efficient}. 
In this scenario the general form of the cost function combines two frequencies: 
\begin{equation}
  E_{VQE} = \kappa\cdot\sin(2p_i+\xi) + \gamma\cdot\sin(p_i+\phi) + C \;. \label{eq:5point}
\end{equation}
The extra sinusoidal term in this expression can be explained by the additional constant term $|c\rangle$ in the quantum state originating from controlled unitary gate operations.
Namely, the state generated by the circuit as a function of the parameter $p_i$ can be written in the form
\begin{equation}
    |\Psi\rangle  = \cos(p_i)|a\rangle + \sin(p_i)|b\rangle + |c\rangle
\end{equation}
By evaluating the expectation value of the Hamiltonian, Eq.~(\ref{eq:5point}) can be recovered.
\pr{(In the appendix we discuss the mathematical details of deriving Eqs.~(\ref{eq:3point}) and (\ref{eq:5point}).)}
Unlike the original setup with $3$ cost function calls, in this case $5$ samples are needed to fully recover the cost function and to solve the minimization problem with respect to a single parameter. 
Computationally, the increased number of the required function evaluations might question the application of this generalized approach, however, the enhanced expressibility of the circuit with parametric controlled rotations might improve the trainability of the circuit, which has been already proven in quantum gate compilation experiments \cite{squander2, Nemkov2023efficient}. 
In VQE experiments controlled rotation gates can be considered as an abstraction, encompassing two parameter-free controlled two-qubit gates and two single-qubit rotations with correlated angles of rotation. 
As was shown in the study of Refs. \cite{Volkoff_2021} and \cite{PRXQuantum.3.010313}, correlated parameters in the circuits might increase their trainability.

\paragraph*{Optimization over gradient-free \pr{batched} search direction:}

To further generalize the optimization approach and potentially improve its efficiency, we extend the concept of the parameter-wise optimization to a finite subset of the parameters. While this extension adds complexity to the line search phase of the optimization, it has the potential to result in much faster convergence of the solution.

\begin{figure}
     \centering
     \includegraphics[width=0.5\textwidth]{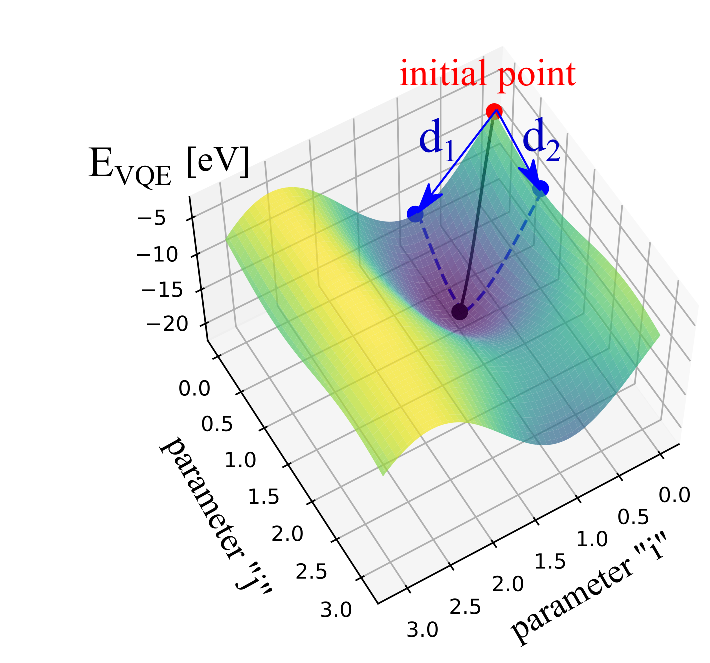} 
     \caption{A typical $E_{VQE}$ energy landscape as a function of two selected parameters "$i$" and "$j$", while the other parameters are kept fixed.
     The blue colored arrows $d_1$ and $d_2$ label the two components of the search vector given by Eq.~(\ref{eq:d}). 
     The search vector determines the path of the line search between the initial (red colored) and minimal point (black colored).
     In the next iteration step a new line search is performed originating from the black point and optimizing over a new subset of parameters.}
     \label{fig:landscape}
 \end{figure}

As mentioned in the preceding section, it requires $3$ function calls to determine the exact minimum of the cost function along a single parameter while keeping the others fixed. However, if the function value at the initial point $\boldsymbol{\theta}$ is known from the previous optimization iteration, the cost of determining the minimum with respect to a single parameter reduces to $2$ function calls. Let's assume that at the start of the iteration step, a subset $\Lambda \subset {1, \dots, L}$ of the parameters was randomly selected (where $L$ is the total number of parameters). For each parameter $\theta_i$ with $i \in \Lambda$, the parameter-wise minimum $\theta_i^*$ was determined using Eq.~(\ref{eq:3point}). 
Then we define the search direction $\mathbf{d}$ in the parameter space with components
\begin{equation}
    d_i = \left\{\begin{matrix}
    \theta_i^* - \theta_i & \textrm{if } i\in\Lambda\\
    0 & \textrm{otherwise.}
    \end{matrix} \right. \label{eq:d}
\end{equation}
The method for determining the gradient-free search direction is illustrated schematically in Fig.~\ref{fig:landscape}. 
In each iteration, the search direction $\mathbf{d}$ is established with respect to the randomly selected subset of parameters. Subsequently, a line search is conducted over the points of $\boldsymbol{\theta} + t\cdot\mathbf{d}$ with $0 \leq t \leq 1$ to identify the point with the lowest cost function, which serves as the starting point for the next iteration.

Following the best practices applied in conventional gradient-based optimization methods, we incorporate an \pr{inexact} line search into the algorithm. 
\pr{The line search process begins by bracketing a local minimum, as described in Ref.~\cite{1606817}. Our algorithm automatically determines the search distance $|\mathbf{d}|$, which establishes the bracketing by providing the starting and ending points of the search interval. The algorithm then estimates the minimum by iteratively narrowing down the search interval, following the approach outlined in Ref.~\cite{1606817}.}

We assert that the described procedure will converge to either a global or local minimum, and this is ensured by two key properties of the algorithm.
(i) Firstly, the cost function consistently decreases with each iteration.
(ii) Secondly, any local minimum in the landscape is an attracting fixed-point of the optimization process.
Therefore, like other optimization strategies, the described algorithm might get trapped in a local minimum as well.
An avenue for significant enhancement could involve executing the optimization with multiple agents and implementing evolutionary selection among them, as detailed in a previous section. However, our numerical experiments have indicated that even with a single agent, the described algorithm proves to be quite efficient in training quantum circuits up to sizes accessible on classical simulators.

\pr{At this point we also remark two subsequent advantages of our \pr{batched} search direction method compared to the conventional gradient based optimization, which can be also added into the described \prr{optimization} framework.
Firstly, when the search direction comes from a gradient-based approach (using either first\cite{https://doi.org/10.48550/arxiv.1412.6980,haug2021optimal,PhysRevResearch.2.043246}  or second order gradient methods\cite{kelley1999iterative}) the range of the accompanied line search must be determined by an additional hyper-parameter of the model.
In contrast, the described gradient-free optimization method inherently evaluates the range of the search direction, without introducing further hyper-parameters in the model.
This intrinsic property \prr{also} makes the optimization process not sensitive to specific parameter configurations which can lead to a vanishing gradient, like in the case of identity-initialization of the circuit to be trained.
Lastly, the outlined method also prevents the optimization process from over-stepping narrow valleys, which is a common weakness of gradient based approaches.}

\section{Discerning BPs via monitoring the entropy of subsystems}

As mentioned in the introduction, a high \prr{value} of entanglement (or von Neumann) entropy \cite{bengtsson_zyczkowski_2006} is often accompanied with the occurrence of BPs in the optimization landscape. 
The rapid \prr{growth} of the entanglement entropy \prr{serves} as an indicator of BP formation, making it a valuable monitoring tool during optimization.
Inspired by these findings, we also use the concept of entropy monitoring during the optimization process to characterize the path in the parameter landscape passed during the optimization process.

Moreover, the study of ref. \cite{PRXQuantum.3.020365} introduces the concept of weak BPs (WBPs) as precursors of BPs.
WBPs emerge when the entanglement of a local subsystem exceeds a certain
threshold identified by the entanglement of a fully scrambled state. 
In contrast to BPs, WBPs can be efficiently diagnosed using the few-body density matrices via evaluating the second-Rényi entropy, given by
\begin{equation}
    S_2 = - \ln \textrm{Tr} \rho_A^2 \label{eq:renyi2}\;.
\end{equation}
Here $\rho_A = \textrm{tr}_B \rho$ is the reduced density matrix where $A$ denotes the subset of qubits that are measured and $B$ is the rest of the system.
While in a simulation the evaluation of the entropy is straightforward, its{its} value can also be inferred experimentally.
The algorithm introduced in \cite{Huang2020}, called classical shadow estimation, efficiently estimates the expectation value not only of the cost function and its gradients but also the second-Rényi entropy of small-sized subsystems.
For practical reasons, we will always normalize the second-Rényi entropy by the Page entropy $S_{Page}$, which is the upper bound of the von Neumann entropy of a Haar random state \cite{PhysRevLett.72.1148}:
\begin{equation}
    S_{Page}(k, N ) \approx k \ln 2 - \frac{1}{2^{N-2k+1}}\;.
\end{equation}
Here $k$ labels the number of qubits in the chosen subspace in the $N$-qubit register.
Study of Ref.\cite{PRXQuantum.3.020365} showed that in the limit of (nearly)
maximal possible entanglement of a small sub-region A, the expected value of the second-Rényi entropy comes close to the Page entropy.

\section{Results}

In this section, we present the \prr{outcome} of our numerical simulations conducted on variational problems, specifically the estimation of the ground state of an XXX Heisenberg Hamiltonian and the SYC model. In both of these models, the ground state satisfies volume law, indicating that the entanglement entropy of any bi-partition of the state scales with the volume, i.e., $(\rho_A) \sim |A|$ (refer to Ref.~\cite{PRXQuantum.3.020365} for a comprehensive review of these concepts).
Additionally, we share numerical results obtained from quantum gate decomposition problem scenarios. Our findings demonstrate that the developed evolutionary algorithm excels in deep quantum circuit synthesis tasks, outperforming traditional optimization methods such as the well-known Broyden–Fletcher–Goldfarb–Shanno (BFGS) \cite{kelley1999iterative} or the ADAM \cite{https://doi.org/10.48550/arxiv.1412.6980} optimization methods. These traditional methods prove ineffective in producing meaningful results for gate decomposition.

We conducted the training of the targeted quantum circuits utilizing the high-performance quantum computer simulator embedded in the SQUANDER package. To assess the capabilities of the new optimization technique, we deliberately excluded any noise in the simulation of the circuits and assumed perfect measurements on the qubits. This approach allowed us to concentrate specifically on the performance of the evolutionary optimizer across various applications and leave the impact of noise and other imperfections of a VQE experiment for future studies.

\begin{figure}
     \centering
     \includegraphics[width=0.8\textwidth]{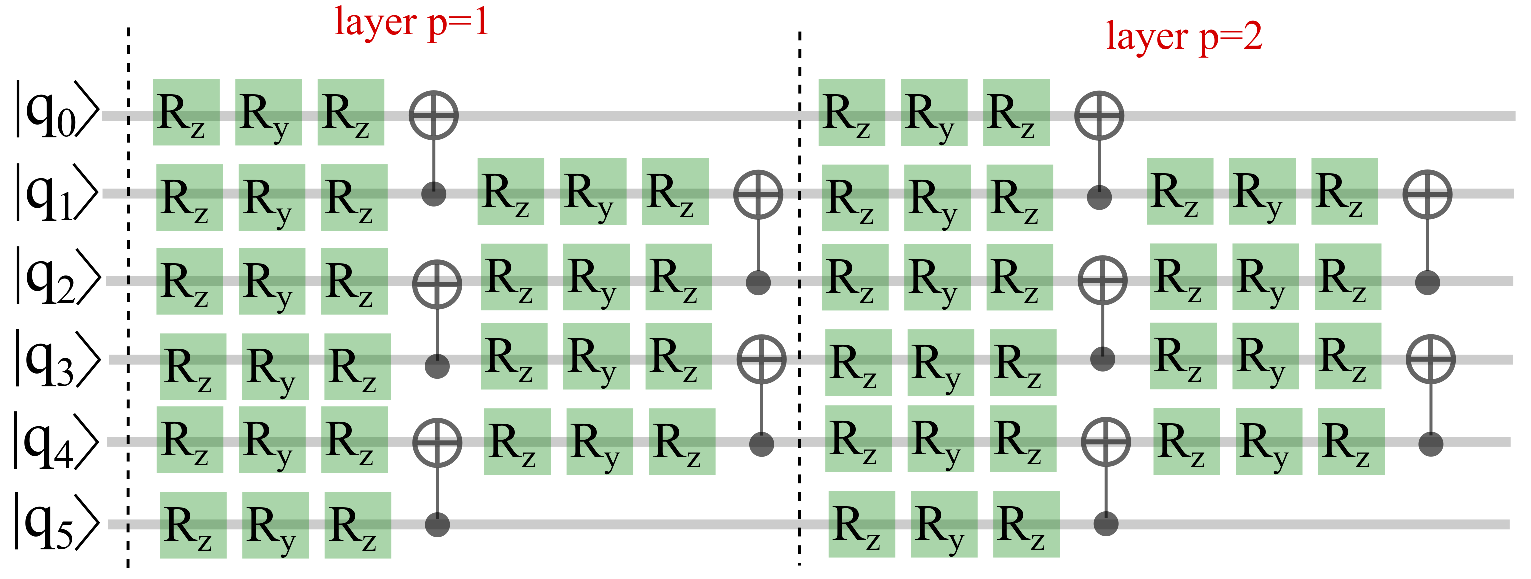} 
     \caption{The structure of the "hardware-efficient" circuit ansatz used in our numerical experiments. The single qubit rotations are expressed by three consecutive rotations around axis $Z$, $Y$ and $Z$  again. The circuit ansatz is built from identical layers, denoted with the parameter $p$.}
     \label{fig:circuit}
 \end{figure}

In the training of quantum circuits, we considered the circuit of the form of "hardware-efficient" ansatz described by a unitary
\begin{equation}
 U\left(\boldsymbol{\theta}\right) = \prod\limits_{i=1}^P\left( W_i \prod\limits_{j=1}^N U_i^j(\theta_i^j)\right)\;.   
\end{equation}
The $N$-qubit circuit is composed from periodic layers of parameter-free entangling layers $W_i$ and parametric single-qubit rotation layers composed from independent single-qubit rotations $U_i^j$ acting on qubit $j$ in the layer $i$. The entangling layers are composed from controlled-not (CNOT) gates connecting nearest neighbor qubits, as shown in Fig.\ref{fig:circuit} for the case of $5$ qubits.

Our initial objective in these numerical experiments is to demonstrate the efficient utilization of the gradient-free \pr{batched} search direction in circumventing BPs during circuit training. To achieve this, we focus on a specific scenario in which only a single agent operates on a randomly chosen subset of the parameters. In this form, the strategy closely resembles traditional gradient descent optimization algorithms.
The key difference is the incorporated routine to determine the search direction.
While conventional gradient-based optimizers choose the search direction based on local properties of the cost-function landscape, such as the gradient or curvature, our proposed alternative approach suggests that determining the search direction via long-range properties of the optimization landscape yields superior results, without {introducing extra hyper-parameters in the optimization strategy}.
Our second objective was to enhance the optimization strategy by introducing mechanisms to prevent or significantly reduce the likelihood of getting trapped in local minima within the optimization landscape. To achieve this, we employed a multi-agent approach and implemented evolutionary selection among them.

We showcase the efficiency of the evolutionary approach on quantum circuit synthesis problems, particularly in scenarios where our previous findings indicated the presence of local minima leading to the termination of most gradient-based optimizer processes. The gate synthesis problem emerged as a promising candidate to investigate the impact of evolutionary selection on the optimization success rate while maintaining a low level of complexity in the cost-function evaluation.
This aspect of our numerical experiments is crucial, as optimization with evolutionary selection involves a significant amount of redundant calculations that need to be discarded.

\paragraph*{Heisenberg XXX Model:}

The Heisenberg model, composed in terms of the $\hat{X}$, $\hat{Y}$ and $\hat{Z}$ Pauli spin operators, encompasses the most general form of Hamiltonian that can be simulated on a qubit-based quantum processor.
When scaled up, the Heisenberg model with randomly chosen coupled qubit pairs, becomes hard to simulate by classical means.
The reason for this is attributed to the non-local interactions between the sites resulting in a volume-law entanglement scaling for a typical bipartite cut.
Specifically, here we focus our study on $k$-local Hamiltonians $\hat{H}$, defined as sum of terms each containing at most $k$ Pauli matrices. 
In our numerical simulations we take $k$ to be finite and fixed, while the number of
qubits $N\gg  k$.
A typical example describing such a model for $k=2$ and including a bias term related to the effect of an external magnetic field is given by equation
\begin{equation}
    \hat{H}_{XXX} = \sum\limits_{i,j\in V_{\mathcal{G}}} J\left( \hat{\sigma}^z_i\hat{\sigma}^z_j + \hat{\sigma}^y_i\hat{\sigma}^y_j + \hat{\sigma}^x_i\hat{\sigma}^x_j\right) + h_z \sum\limits_N \hat{\sigma}^z_i \;.
\end{equation}
Despite of the local $k$-terms in the formula, according to the reasoning of Refs.~\cite{McClean2018,Cerezo2021} the VQE formulation of such models will still exhibit BPs in a limit of $\mathcal{O}(poly(N))$ deep quantum circuits.
Therefore, with a deep enough quantum circuit the VQE algorithm will also
suffer from a BP, making this model to be a suitable test case to show the efficiency of the developed evolutionary optimization technique in training the circuit.
\begin{figure}
     \centering
     \includegraphics[width=0.8\textwidth]{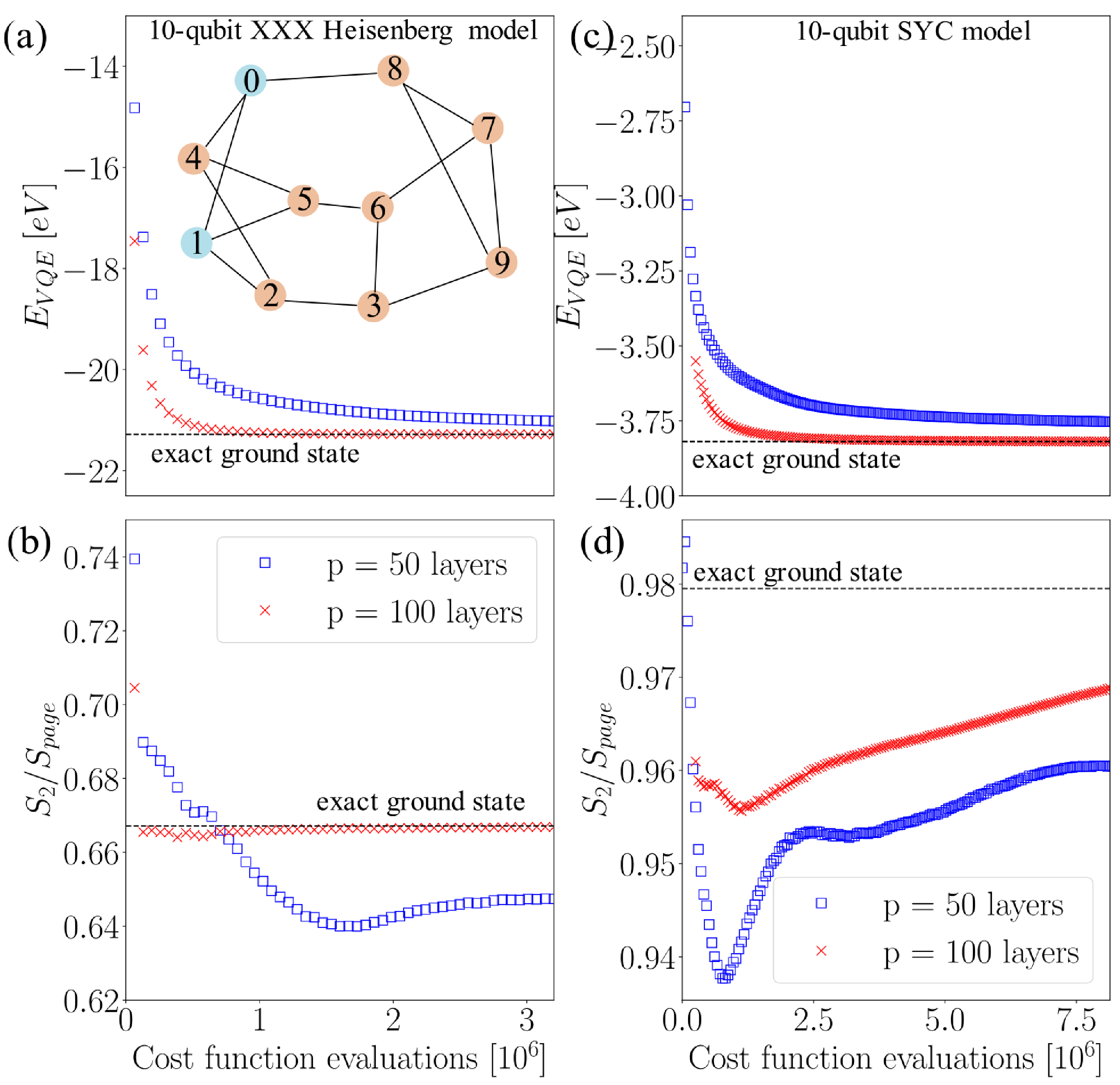} 
     \caption{VQE simulations implementing the strategy based on the gradient-free \pr{batched} search direction with a single agent with $L=128$.
     The training of the quantum circuits was conducted to find the ground state for the $N=10$-qubit XXX Heisenberg model on a $3$-regular random graph depicted in the subset of sub-figure (a) and for a $10$-qubit SYC model.  
     Panels (a) and (b) show the evolution of the VQE energy and the normalized second-Rényi entropy evaluated on the sites $0$ and $1$ (labeled by blue in the insets) for the Heisenberg model.
     Sub-figures (c) and (d) show the same kind of results for the SYC model. 
     The energy and the normalized entropy of the exact ground state for the Heisenberg system are $E_0=-21.28$ eV and $S_2/S_{Page} = 0.667$.
     The exact ground state of the SYC model is $E_0=-3.82$ eV with normalized entropy $S_2/S_{Page} = 0.980$.
     For the initialization we used random parameters.}
     \label{fig:10qubits}
 \end{figure}

\paragraph*{Sachdev–Ye–Kitaev model:}

From theoretical point of view, the Sachdev–Ye–Kitaev (SYK) model- \cite{SYC,PhysRevLett.70.3339,PhysRevD.94.106002}, with it's non-local Hamiltonian structure in the Puali basis, is a frequently visited representative of variational problems that are proven to be hard to solve in VQE context.
This is because the SYC model features a ground state that is nearly maximally entangled \cite{PhysRevD.100.041901,PhysRevD.94.106002}, so crossing a BP becomes inevitable during the VQE process \cite{Kim_2022}.
From solid state physics point of view, the SYK model can be used to describe $2N$
spinless Majorana fermions $\chi_i$ satisfying the following anti-commutation relations $\{\chi_i, \chi_j\} = \delta_{i,j}$. 
The Hamiltonian of the model reads as
\begin{equation}
    \hat{H}_{SYC} = \sum\limits_{1\leq i < j < k <l\leq 2N} J_{i,j,k,l}\;\chi_i \chi_j \chi_k \chi_l\;,
\end{equation}
where the couplings $J_{i,j,k,l}$ are taken randomly from a Gaussian distribution with zero mean and a variance of
\begin{equation}
    \textrm{var}\left(J_{i,j,k,l}\right) = \frac{3!}{(N-3)(N-2)(N-1)}J^2\;.
\end{equation}
This fermionic Hamiltonian can be adopted to a quantum hardware by mapping the couplings between the Majorana fermions onto spin-chain variables by the nonlocal Jordan-Wigner transformation:
\begin{equation}
    \chi_{2i} = \sigma^x_1\sigma^x_2\dots \sigma^x_{i-1}\sigma^y_i \qquad
    \chi_{2i-1} = \sigma^x_1\sigma^x_2\dots \sigma^x_{i-1}\sigma^z_i\;.
\end{equation}
such that $\{\chi_i, \chi_j\} = \delta_{i,j}$ will hold on. 
This encoding enables one to model $2N$ Majorana fermions with only $N$ logical qubits. 
Generally speaking, we can consider the SYC model as a corner case of the Heisenberg model, when no restriction is made on the number of coupled qubits in a single term in the Hamiltonian.
For our studies, we set $J = 1$ and consider a system of $N = 10$ qubits.

\paragraph*{Numerical results:}

\pr{Firstly, in Fig.~\ref{fig:10qubits} we present the results of our numerical simulations conducted on a $10$-qubit Heisenberg XXX and SYC models.
We wanted to address the same physical systems as in prior study of Ref.~\cite{PRXQuantum.3.020365} giving us the opportunity to compare our method to what was described in \cite{PRXQuantum.3.020365}.
The strategy developed in \cite{PRXQuantum.3.020365} work relies on the monitoring of the second-Rényi entropy (\ref{eq:renyi2}) and reducing the learning rate when the entropy builds on high values.
}

\pr{Though, our numerical results revealed that the addressed system is still too small without the requirement of any special treatments to mitigate the BP effects, provided the VQE energy can be obtained with ideal measurement.
Notably, our approach (and other gradient-based strategies) is capable to solve the optimization problem without the need for external interventions or monitoring of specific quantities:
during the calculations the second-Rényi entropy autonomously remained at low values. 
}

The left side of Fig.\ref{fig:10qubits} illustrates our VQE experiments on a $10$-qubit Heisenberg model identical to the system studied in \cite{PRXQuantum.3.020365}. In each iteration step, we performed optimization by randomly selecting $L=64$ parameters of the circuit and conducted a line search along the direction determined by Eq.(\ref{eq:d}). Throughout the iterations, we utilized a single agent, and no evolutionary steps were applied during the simulations. Still, the optimization problem could be solved, while significantly reducing the computational complexity.

The outcome of this optimization process displayed a steady decrease in the VQE energy during the iterations [see Fig.\ref{fig:10qubits}.(a)], while the evolution of the second-Rényi entropy demonstrated a rather non-trivial pattern [see Fig.\ref{fig:10qubits}.(b)]. As the VQE instances in our simulations were initiated to random parameters, the entropy at the beginning of the optimization picked up large values implying large entanglement in the resulting state vector.
However, as the optimization progressed, the entropy rapidly decreased, even falling below the value calculated from the exact ground state. Given that a sharp decrease in the evolution of the cost function generally correlates with low values of the second-Rényi entropy, this observation is expected. As the search direction and the bounds of the line search are determined via globally-scaled properties of the optimization landscape, the optimization process can navigate away from flat areas and avoid overstepping narrow valleys, guiding toward acceptable solutions.

Upon further analysis of the numerical results, it is valuable to explore the resulting properties of the quantum state $|\Psi(\boldsymbol{\theta})\rangle$ obtained via circuits of varying depths. In addition to examining the energy and entropy associated with the resulting state, a direct comparison between the state $|\Psi(\boldsymbol{\theta})\rangle$  and the exact ground state can be characterized by their overlap integral
\begin{equation}
    M =  \left|\langle\Psi_0  | \Psi(\boldsymbol{\theta}) \rangle\right|^2  \label{eq:overlap}
\end{equation}
Values of $M$ close to $1$ signify that the VQE wave function faithfully reproduces the exact ground state. On the left side of Fig.\ref{fig:10qubits}, we compare a VQE optimization performed with circuits of depths $p=50$ and $100$ layers (refer to Fig.\ref{fig:circuit} for the circuit structure).
As depicted, with $100$ layers, both the VQE energy and the second-Rényi entropy closely approach the theoretical results of the exact ground state. The overlap integral $M$ calculated for the produced VQE state is also very close to $1$, indicating that the VQE experiment successfully reproduced the exact ground state with high accuracy. (The resulting VQE circuit with parameters can be accessed within the SQUANDER package \cite{SQUANDER_github}).

For a circuit with $50$ layers, on the other hand, we could not approximate the ground state with such high quality. Neither the energy nor the entropy [plotted in blue in Fig.~\ref{fig:10qubits}.(a)-(c)] fully converged to the values corresponding to the exact ground state. The overlap integral at the end of the optimization (terminated at $\sim 3\times 10^6$ cost function calls) is $M=0.97$. The results obtained for even shallower circuits rapidly become inaccurate. The explanation behind this observation is related to the expressiveness of the circuits. Considering $10$ qubits, a general wave-function can be described by $2^{10}=1024$ complex amplitudes, resulting in $2\times2^{10} -2 = 2046$ free parameters (two parameters are constrained since the state is normalized). The $50$-layered VQE circuit has $2700$ free parameters, while the $100$-layered variant can be tuned with $5400$ free parameters. As one can see, the $50$-layered circuits contain enough parameters to encode any quantum state, but the convergence slows down at the end of the optimization. The $100$-layered circuit converges much faster to the solution, as it is likely over-parameterized.

Our results obtained for the SYC model further validate these considerations. As depicted in the right panels of Fig.~\ref{fig:10qubits}, the VQE optimization for the SYC model exhibits similar behavior to the Heisenberg model on the left. With $p=100$ layers in the circuit ansatz, we achieved a decent approximation of the ground state energy. In contrast, at $p=50$ layers, the approximation becomes less accurate after $6.5\times10^{-6}$ evaluations of the cost function. The second-Rényi entropy remains high throughout the optimization process, though it does not hinder the optimizer from obtaining high-quality results.

Regarding the exact solution of the SYC model, we need to mention that the diagonalized SYC Hamiltonian revealed a $2$-fold degeneracy of the states. In addition, the $2$-fold ground state and the nearest excited states (also $2$-fold degenerate) were close to each other in energy. While the energy of the ground state was $E_{0,SYC}=-3.819$ eV, the first excited energy was $E_{1,SYC}=-3.799$ eV. The VQE energy obtained with the training of the $p=100$ layered circuit turned out to be $E_{VQE}=-3.814$ eV, a value that falls between the ground state energy and the first excited energy. Thus, the VQE state is expected to have a finite component orthogonal to the subspace of the $2$-fold ground state. Fortunately, this orthogonal component has only an overlap $M=0.098$ with the subspace of the first excited states, while the overlap of the VQE wave function with the subspace of the $2$-fold ground state is $M=0.902$. Hence, the components of the VQE wave function predominantly come from the ground state and the first excited state.

\begin{figure}
     \centering
     \includegraphics[width=0.8\textwidth]{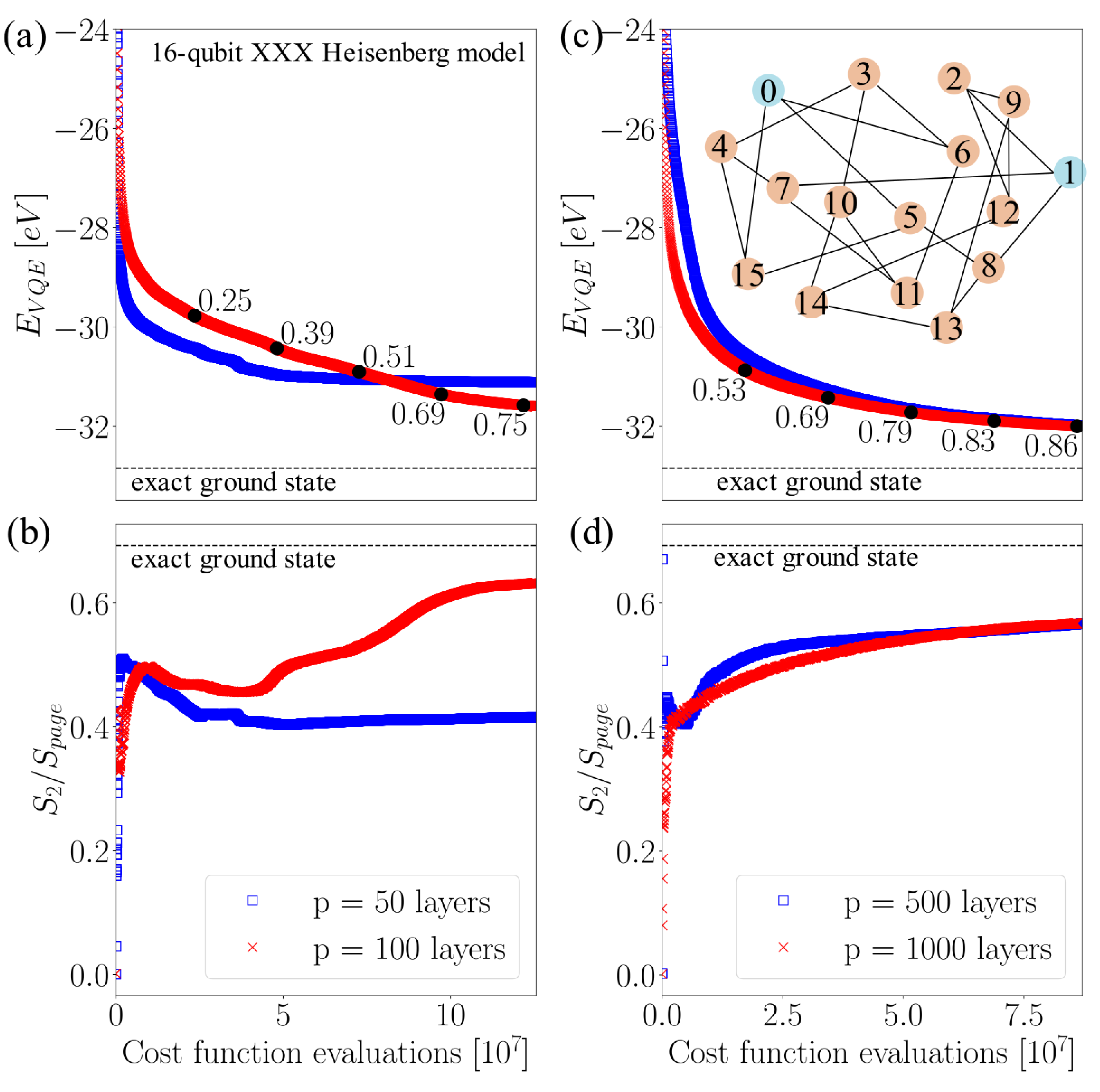} 
     \caption{VQE simulations implementing the strategy based on the gradient-free \pr{batched} search direction with a single agent with $L=128$.
     The training of the quantum circuits was conducted to find the ground state for the $N=16$-qubit XXX Heisenberg model on a $3$-regular random graph depicted in the subset of subfigure (c).  
     Panels (a) and (b) show the evolution of the VQE energy and the normalized second-Rényi entropy on sites $0$ and $1$ (labeled by blue in the insets) for the Heisenberg model evaluated with shallower circuits consisting of $p=50$ and $100$ layers, while .
     Subfigures (c) and (d) show the same kind of results obtained with deep circuits consisting of $p=500$ and $1000$ layers. 
     The black numbers and dots represent the evolution of the of the overlap integral between the VQE wave function and the exact ground state defined by Eq.(\ref{eq:overlap}).
     The energy and the normalized entropy of the exact ground state for the Heisenberg system are $E_0=-32.84$ eV and $S_2/S_{Page} = 0.693$.
     The optimization was initialized with parameters equal to $0$.}
     \label{fig:16qubits}
 \end{figure}

After examining $10$-qubit systems, we raise the stakes and turn our attention to even more challenging problems. Although it is not a significant challenge for quantum computer simulators to execute circuits beyond $20$ qubits, VQE experiments become quite expensive at this scale due to the large number of cost function evaluations. Therefore, in this work, we limit our considerations to problems up to \prr{$21$} qubits.
\prr{At this point} the goal of our VQE experiments is to reproduce the ground state of a randomly generated $16$-qubit XXX Heisenberg model (with $k=3$), as visualized in the inset of Fig.~\ref{fig:16qubits}.(c). Furthermore, to increase the efficiency of the optimization process, we initiate the training of the circuits with parameter values equal to $0$. Since the circuits are applied to the initial state $|\mathbf{0}\rangle$ and initially, all single-qubit rotations will be identities with this choice, the two-qubit entangling gates will lose their effect as well. This is because all control qubits are initially in state $0$, so the controlled two-qubit gates incorporated into the circuit will act as an identity gate as well. Consequently, the resulting state will have zero initial entanglement entropy, promising a steep initial descent in the VQE energy.
\prr{During the optimization}, our method remains autonomous in avoiding BPs on its own, there is still no need for any external intervention during the optimization.

A general $16$-qubit wave function can be characterized by $2^{16}-2=131070$ parameters. Consequently, circuits with either $p=50$ or $p=100$ layers are expected to provide much less accurate VQE wave functions than in the previously addressed $10$-qubit case, where the number of trainable parameters in the circuit was sufficiently large. In the $16$-qubit case, the $100$-layered circuit ansatz contains only $9000$ free parameters, which is much less than required to approximate the target state.
The left panels of Fig.\ref{fig:16qubits} show the numerical results of the conducted VQE experiments with circuits consisting of $p=50$ and $p=100$ layers. As expected, it was not possible to precisely approximate the ground state energy in these instances (VQE energy of $-31.1$ eV with the $50$-layered circuit and $-31.6$ eV with the $100$-layered circuit). The $100$-layered circuit gave significantly better results than those achieved with the $50$-layered circuit. However, neither of these circuits could get close to the exact ground state energy of the model. After more than $10^8$ evaluations of the cost function, the VQE wave function still showed a significant difference compared to the exact ground state, and the overlap calculated with the wave function generated by the $100$-layered circuit did not exceed the value of $M=0.75$. The black numbers in Fig.\ref{fig:16qubits}.(a) indicate the evolution of the overlap with the number of cost function evaluations.
As mentioned, these findings were not surprising, as the number of trainable parameters in these circuits was only $4500$ and $9000$ for the $50$- and $100$-layered circuits, respectively. The expressiveness of these circuits lags behind the numbers needed for the accurate approximation of a general $16$-qubit wave function. Therefore, we increased the number of layers in our circuit ansatz by an order.

The right panels of Fig.\ref{fig:16qubits} show our numerical results obtained with these deep circuits composed of $p=500$ and $p=1000$ layers. As the plotted results indicate, the developed optimization strategy had no difficulties proceeding with the training of these deep circuits. We still executed only a single agent in these numerical experiments, and optimized $L=128$ parameters in each iteration. We observed that the efficiency of the optimization varies from run to run. Fig.\ref{fig:16qubits} shows the most successful executions, while in several experiments, we observed signs of convergence to a local minimum around the energy level of $\sim -29$ eV. We argue that this issue could be resolved by increasing the number of agents in the process due to the evolutionary selection of the most successful agents. However, due to the overwhelming numerical requirements of the performed VQE simulations, we did not try out this approach in these simulations. We revisit this aspect of the algorithm in a later section by addressing quantum gate synthesis problems, where it becomes essential to avoid local minima to retrieve meaningful results. In these experiments, evolutionary selection will play a crucial role in solving the addressed problems.

In the conducted $16$-qubit VQE optimization our numerical simulation yielded a VQE energy of $E_{VQE}=-31.99$ eV with $p=500$ layers and $E_{VQE}=-32.00$ eV with $p=1000$ layers after $8\times10^{7}$ cost function evaluation, at which point the converge towards the exact ground state energy has dramatically slowed down. 
The largest value of the overlap integral (\ref{eq:overlap}) with the exact ground state turned to be $M=0.86$.

\paragraph*{Impact of qubit count on the optimization performance:}

\prr{In the previous section, we demonstrated our optimization technique on a $16$-qubit system. The quantum circuit used in that setup was designed with a layer count that yielded a number of tunable parameters comparable to the degrees of freedom of a $16$-qubit quantum state, enabling a close approximation of the target ground state (see the numbers indicating the overlap integrals in Fig.\ref{fig:16qubits}. However, as the number of qubits increases, maintaining this correspondence would require an exponential increase in circuit depth. This poses a dual challenge: not only does the expressivity requirement grow exponentially, but also the computational costs of quantum circuit emulation, even if the number of layers is kept constant. To partially address this issue and explore the scalability of our optimization method, we present numerical results for VQE problems with qubit counts ranging from $17$ to $21$, using circuits with a fixed depth of $L=500$ layers. Further investigations of our method’s behavior and performance reported in subsequent sections continue to focus on the $16$-qubit regime, enabling us to effectively approximate the ground state of the targeted system.}
\begin{figure}
     \centering
     \includegraphics[width=0.5\textwidth]{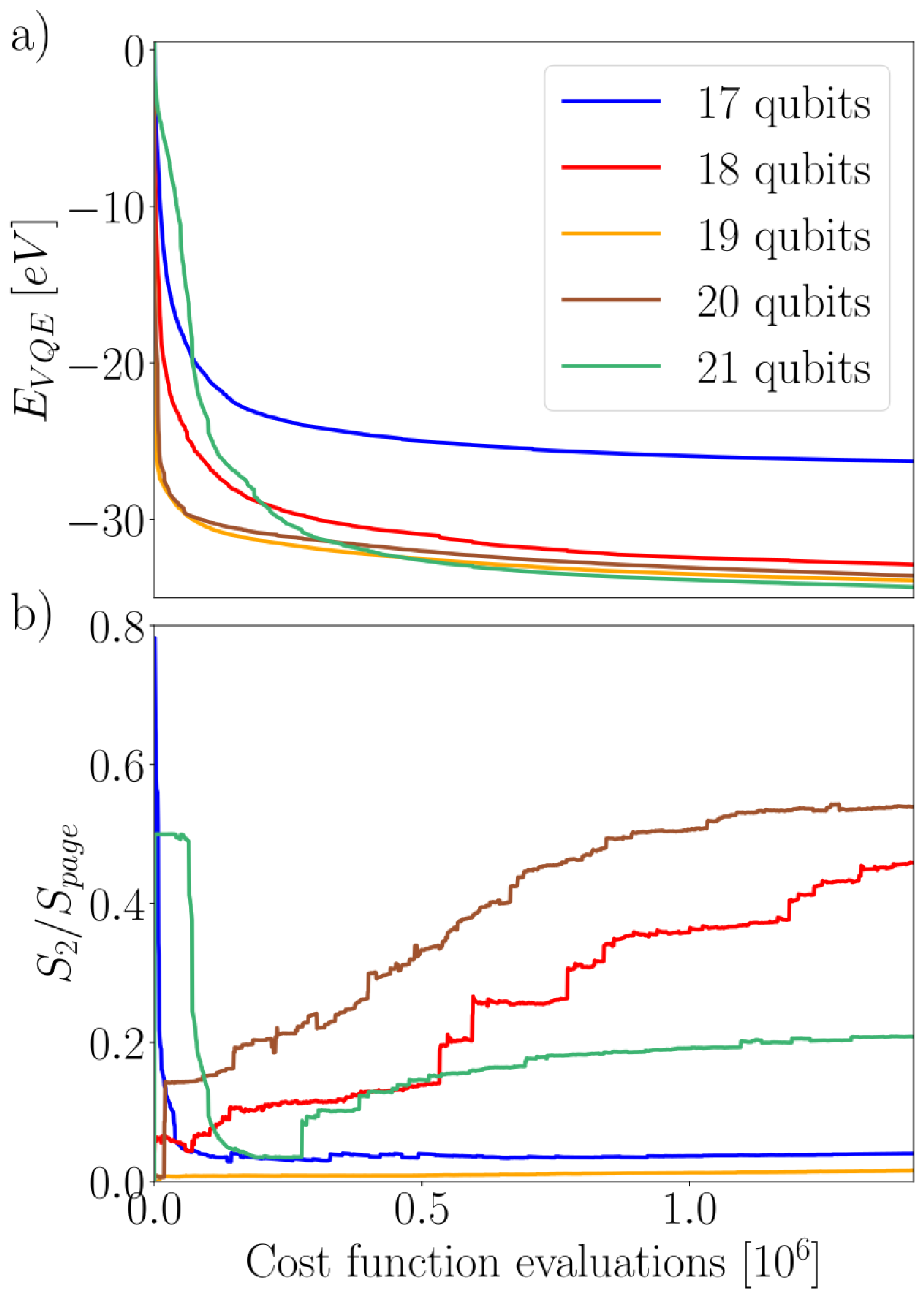} 
     \caption{\prr{(a) Optimization processes to solve Eq.(\ref{eq:EVQE}) on randomly generated $17$, $18$, $19$, $20$ and $21$-qubit Heisenberg models. The entanglement entropy shown in subfigure (b) stayed at a low value for all of the addressed qubit counts.
     We executed the optimization method developed in this work with $L=32$.}}
     \label{fig:scaling}
 \end{figure}
 \prr{Figure \ref{fig:scaling} illustrates that our optimization method, when applied to randomly generated Heisenberg systems, consistently avoided BPs throughout the optimization process. This is supported by the low entanglement entropy values observed across all examined cases. After a sharp initial drop in the VQE energy during the first $7\times10^5$ cost function evaluations, the optimization curves begin to level off, suggesting convergence toward a local minimum.
This observation is further supported by analyzing the overlap integrals computed with the exact eigenstates of the addressed physical systems: while the overlap integrals 
form a shallow distribution taking on value between $10^{-6}$ and $10^{-3}$, one or two values calculated with low-energy eigenstates start to emerge from this distribution.
Getting trapped in local minima is a common issue of optimizers.
Later in this study we will propose an evolutionary algorithm providing an efficient strategy to escape from them via evolutionary selection rules. 
We will demonstrate the efficiency of this approach on smaller problems formed for quantum circuit compilation.
}

\prr{
These results reinforce the assumption that the optimization process tends to converge toward a local rather than a global minimum. However, due to the lengthy runtime of the optimizer, the convergence was intentionally stopped at an early stage. This decision was driven by the primary objective of our numerical experiments: not to find the final solution, but to evaluate the robustness of the optimization process in the presence of BPs in the optimization landscape.

To make the reported experiments feasible within a practical timeframe, we employed hardware acceleration to significantly speed up quantum circuit emulation. Specifically, we developed a state vector simulator implemented on Groq's data-flow hardware architecture\cite{10.1109/ISCA45697.2020.00023}, which we then integrated into the optimization framework used in this study.
This integration resulted in up to a $40\times$ speedup in quantum circuit emulation compared to using a 32-core AMD EPYC 7542 processor, within the qubit range considered in our numerical experiments. This performance gain stems from the fact that parallel evaluation of quantum circuits at this scale does not fully utilize the capabilities of modern CPUs. Similarly, GPUs offer limited advantage in this regime, as confirmed by benchmark data provided by QuantumAI’s emulation service \cite{openAI}.
Moreover, our optimization method does not naturally support parallelization, particularly during the line search phase, where cost function evaluations are inherently sequential. As a result, GPU resources cannot be efficiently saturated by the workload required for our experiments.
In contrast, Groq's Language Processing Units are particularly well-suited for this task, as they are optimized for efficient moving of data over the chip enabling fast linear transformations on state vector element pairs. This capability allowed us to significantly accelerate the optimization process, enabling us to report on the results presented in Fig.~\ref{fig:scaling}.
}

\paragraph*{Comparison to conventional gradient descent and gradient free methods:}

\begin{figure}
     \centering
     \includegraphics[width=0.8\textwidth]{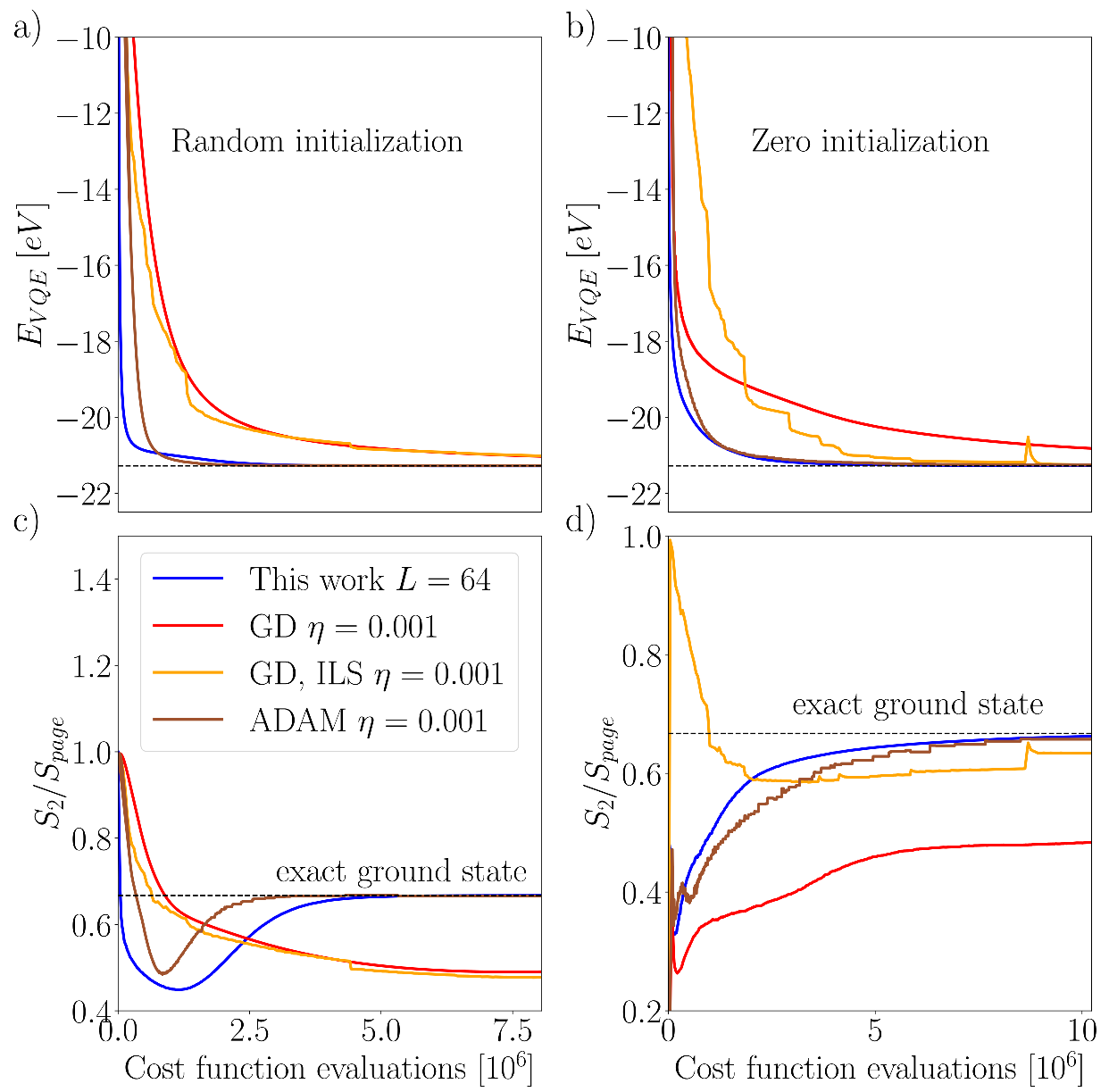} 
     \caption{\pr{Performance comparison of $4$ optimization strategies to solve Eq.(\ref{eq:EVQE}) on a $10$-qubit Heisenberg model identical to the system plotted in Fig.\ref{fig:10qubits}. The blue colored line stands for the optimization method developed by this work with $L=64$. 
     Red lines correspond to basic gradient descent optimization with a learning rate $\eta=0.001$.
     Orange lines represent results obtained with a gradient descent method (GS) with inexact line search (ILS) and the same learning rate parameter. 
     Brown colored lines show the results obtained with the ADAM optimization method with a learning rate $\eta$.}}
     \label{fig:10qubits_comparison}
 \end{figure}

\pr{To assess the computational efficiency of our novel optimization method, we conduct a comparative analysis with traditional gradient descent (GD) approaches. We begin by evaluating the optimization capabilities of each method on a smaller system, and then scale up to a larger system to examine how their efficiency holds up under increased complexity.
On the lower end of the range we refer to our results plotted in Fig.~\ref{fig:10qubits_comparison}.}
\pr{To address the optimization problem (\ref{eq:EVQE}), we explored four strategies available within the SQUANDER package. The Powell-variant of the BFGS optimizer \cite{Powell1987} and the GD methods incorporate an inexact line search\cite{Nocedal2006} in each iteration, proving to be quite efficient for smaller problems (e.g., in the synthesis of $3-5$ qubit circuits). Though, the BFGS method, as a second order method is loosing the performance at large parameter count, hence we omit this method from our comparison on VQE experiments.
The handcrafted implementations of these optimizers also consider the periodicity of the optimization landscape by appropriately bounding the line search during optimization.
Additionally, we included the ADAM optimization algorithm in our tests, although we did not implement the adaptive learning rate strategy proposed in \cite{PRXQuantum.3.020365}.
In Fig.\ref{fig:10qubits_comparison}. we see that the examined optimization engines \prr{have} similar performance in solving the optimization problem, with the best performance attributed to the ADAM method and to the strategy developed by this work.
These two optimizers had much bigger steepness in the cost function curve for both zero and random initializations.
The dominant part of the optimization happened in the first quarter of the performed cost function evaluations, while further refinement of the results got significantly slowed down. 
The steep decrement in the cost function at the beginning can be translated into a reduction in the number of required cost function evaluations to achieve reasonable result, possibly \prr{making} faster the optimization over BPs.
\begin{figure}
     \centering
     \includegraphics[width=0.8\textwidth]{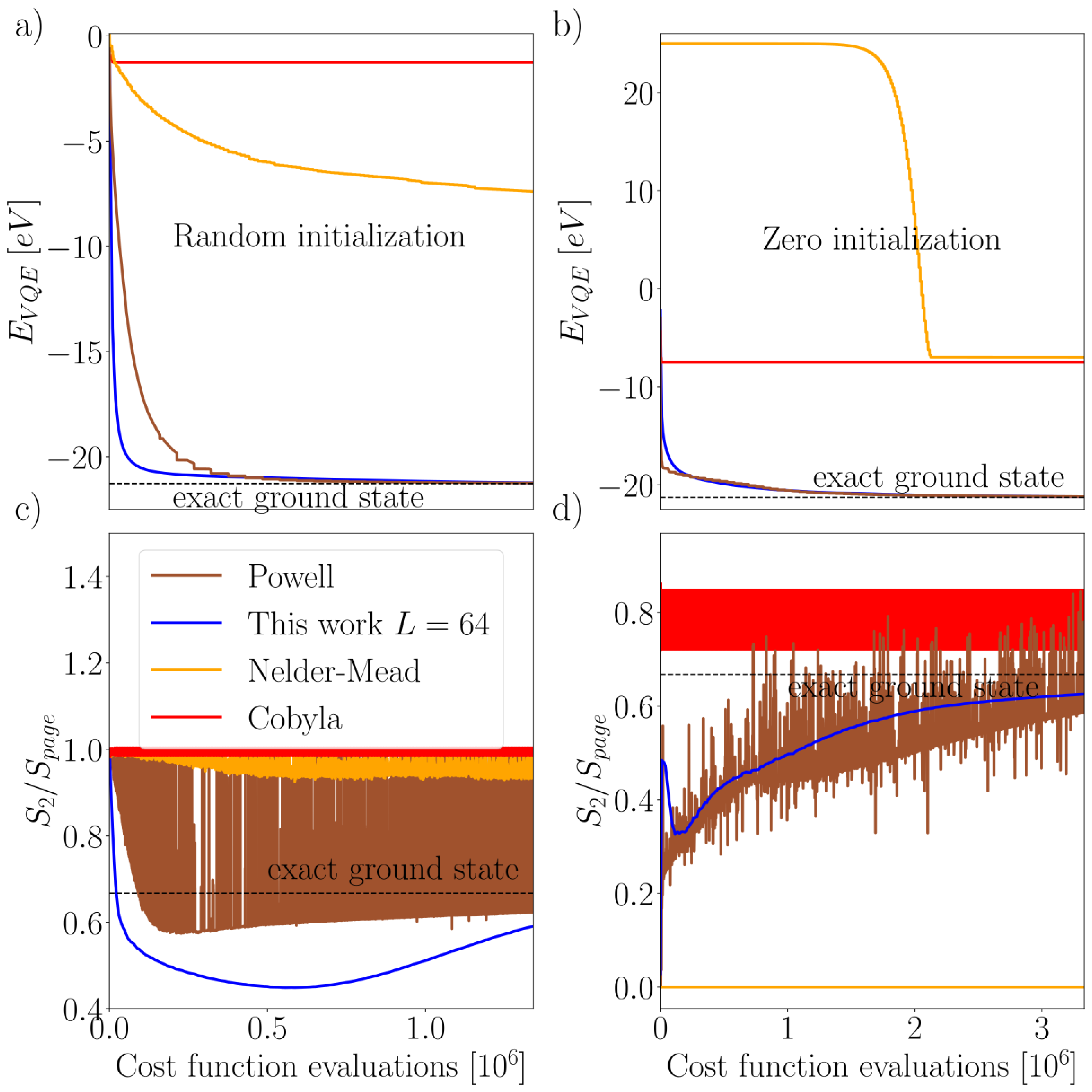} 
     \caption{\pr{Performance comparison of $4$ gradient free optimization strategies to solve Eq.(\ref{eq:EVQE}) on a $10$-qubit Heisenberg model identical to the system plotted in Fig.\ref{fig:10qubits}. The blue colored line stands for the optimization method developed by this work with $L=64$. 
     Red lines correspond to the Cobyla method, orange lines represent results obtained with the Nelder-Mead optimization and the brown line stands for the results obtained with the Powel gradient free method. 
     The gradient free optimizers are taken fro the SciPy package.}}
     \label{fig:10qubits_comparison_grad_free}
 \end{figure}
As our newly developed method reminds the basic concepts of derivative-free approaches, we also explore the capabilities of gradient-free algorithms for comparison. (We took the implementations of the explored gradient-free optimization methods from the SciPy library.) However, we don't consider our method as a gradient-free approach because our method reproduces the whole cross section of the optimization landscape along single parameter directions. The information on the gradient components also can be accessed with no further computation costs. 

In Fig.~\ref{fig:10qubits_comparison_grad_free}. we show the results of our numerical simulations on the same $10$-qubit system as before. We found that two gradient-free optimizers, Nelder-Mead\cite{10.1093/comjnl/7.4.308} and Cobyla\cite{Powell1994}, were unable to solve the $10$-qubit optimization problem. This is in contrast to a prior study\cite{Arrasmith2021effectofbarren} reporting adequate capability for these optimizers in solving similar problems. However, there are key differences between the two studies. For example, Ref.~\cite{Arrasmith2021effectofbarren} used a limited number of parametric layers, whereas our study used a larger number of layers in the circuit ansatz.
Secondly, the objective of Ref.\cite{Arrasmith2021effectofbarren} was to reproduce the identity matrix from a circuit instantiated with randomly generated parameters.
Thus, the target of the optimization in Ref.~\cite{Arrasmith2021effectofbarren} is presumably not located on a BP either.

In contrast, Powell's gradient free optimization method\cite{10.1093/comjnl/7.2.155} showed surprisingly good efficiency in training the addressed quantum circuits.
In fact, Powell's method shares many similarities with our approach, specifically in its manner of use of line searches. 
The method begins by defining an initial set of search directions, typically coordinate directions in the parameter space.
It then iteratively searches along each of these directions, seeking the optimal displacement that minimizes the cost function while only adjusting the parameters along that specific direction.
This process typically employs a variant of the line search algorithm to determine the optimal displacements.
After exploring all the search directions, the method concludes the iteration by replacing the search direction corresponding to the largest displacement with a linear combination of the original search directions.
This process is repeated iteratively during the optimization, with the iteration terminating when a convergence criterion is met.

However, our method has some advantages over Powell's method. For example, our method can determine the direction of the search direction in parallel, which allows for concurrent executions and reduces the execution time. Additionally, our method takes advantage of the properties of the optimization landscape to determine the maximal distance of the line search.

Powell's method was able to solve the optimization problem efficiently, but it exhibited abrupt oscillations in the second-Rényi entropy, which may indicate a loss of efficiency on larger problems. Since the addressed optimization problem is relatively small, it's too early to draw conclusions about the efficiency of the best-performing optimization strategies at scale.
We tried to fill up this gap via further numerical experiments performed on larger systems.
Our numerical results showed that by increasing the system size the efficiency of most of the optimization engines \prr{starts} to drop radically.} 
\pr{We showcase this tendency via numerical experiments performed on the $16$-qubit Heisenberg model depicted in Fig.\ref{fig:16qubits} with $p=500$ circuit layers.
This model delivers high degree of computational complexity for \prr{exhaustive} numerical investigations.}
\begin{figure}
     \centering
     \includegraphics[width=0.8\textwidth]{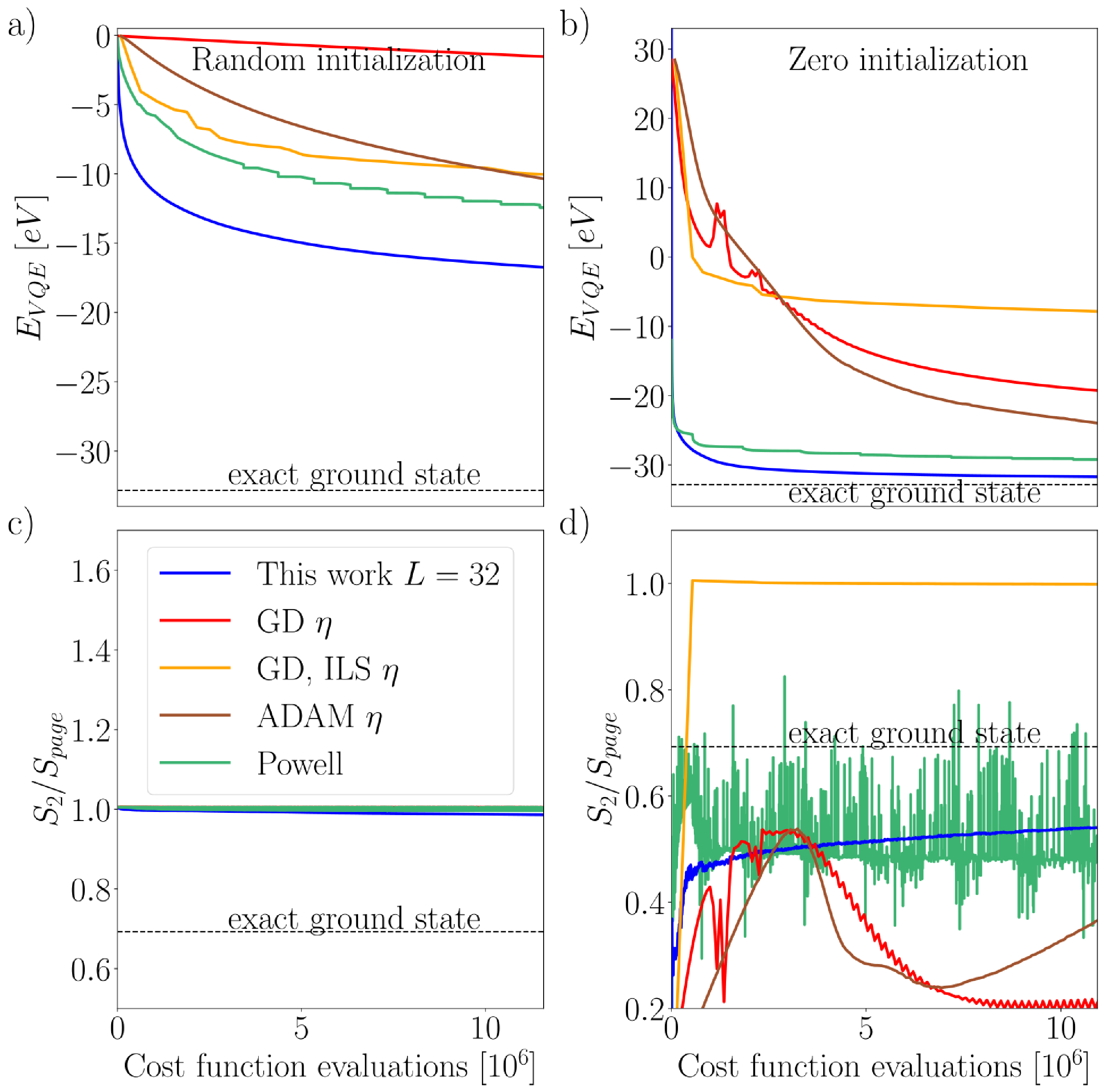} 
     \caption{\pr{Performance comparison of $4$ optimization strategies to solve Eq.(\ref{eq:EVQE}) on a $16$-qubit Heisenberg model identical to the system plotted in Fig.\ref{fig:10qubits}. The blue colored line stands for the optimization method developed by this work with $L=64$. 
     Red lines correspond to basic gradient descent optimization with a learning rate $\eta=0.0001$.
     Orange lines represent results obtained with a gradient descent method (GS) with inexact line search (ILS) and the same learning rate parameter. 
     Brown colored lines show the results obtained with the ADAM optimization method with a learning rate $\eta$. Finally, green color stands for the results of Powell's gradient-free optimizer.}}
     \label{fig:16qubits_comparison}
 \end{figure}
\pr{Our analysis of the larger system reveals a significant difference in optimization performance when starting from random versus (near-)zero initial parameter vectors. When initialized randomly, all five examined optimizers struggled to solve the optimization problem [as shown in Fig.\ref{fig:16qubits_comparison}.a)]. This is because the optimizations began at a BP, making it difficult to approach the global minimum of the VQE energy.
The second-Rényi entropy remained almost constant throughout the iterations [Fig.\ref{fig:16qubits_comparison}.c)], indicating that the trajectories did not escape the BP. As a result, the VQE energies improved only marginally as the number of cost function evaluations increased. Notably, the method developed in this work still demonstrated the best performance in solving the optimization problem.}

\pr{When initializing the optimization with identity operators (i.e., starting with zero rotational parameters), the second-Rényi entropy begins to evolve from a small value, potentially enabling a steeper decrease in the cost function [as shown in Fig.~\ref{fig:16qubits_comparison}.d)]. Still, our comparison reveals that gradient-based optimizers exhibit significantly lower efficiency in solving the $16$-qubit problem compared to the $10$-qubit VQE problem.
This inefficiency is likely linked to the second-Rényi entropy, as the ADAM and GD optimizers, which build up small values of the second-Rényi entropy, are the most efficient among the gradient-based optimizers. In contrast, the gradient descent method with inexact line search enters a region of high entropy early in the optimization, making it highly inefficient.
In contrast, the optimization method developed in this work maintains its efficiency observed for the $10$-qubit system during the first quarter of the iterations, rapidly approaching the target value of the optimization [as shown by the blue line in Fig.~\ref{fig:16qubits_comparison}.b)]. However, when the second-Rényi entropy builds up higher values, the steepness of the VQE energy significantly reduces, making only a marginal improvement on the VQE energy.
Unfortunately, our numerical experiments were unable to fully recover the target value of the exact ground state. As discussed earlier, the used circuit ansatz may not be optimized for the addressed physical system, potentially lacking the necessary expressiveness to capture the ground state. We leave the study of possible impacts of various circuit structures on the quality of the converged solution for future work.
Powell's method also showed outstanding efficiency in decreasing the cost function, it performed almost equally well as our optimization method. Nevertheless, Powell's method has a significant limitation. Due to its sequential nature, it becomes impractical for large-scale problems, making it less suitable for VQE tasks at scale.
Also, we should notice another limiting well pronounced steps (see the green lines in Fig.~\ref{fig:16qubits_comparison}.a) and b)). 
In Powell's method the search directions are determined iteratively from the search directions of the previous step and therefore the optimization along the search directions will affect more and more parameters at once as the optimization iterations go on.  
Due to our reasoning this is a key component in limiting the efficiency of an optimization method to solve the VQE problem at scale.
To overcome this issue, we can restart the solver's inner state and start over the optimization with search direction aligning with single parameter directions as we do in the optimization method developed in this work. The steps in the green curves in Fig.~\ref{fig:16qubits_comparison}.a) and b) indicates the resets of Powell's optimization method. This approach allows us to lift the limitation and continue the improvement on the cost function. 
The other two gradient free optimizers used in our $10$-qubit simulations did not \prr{performe} well there, hence we omitted them from our larger experiments on $16$ qubits.}

\pr{To understand the underlying factors contributing to the success of the optimization method presented in this work, we will analyze its performance and compare it to that of gradient-based optimizers. Specifically, we will investigate why our method outperforms the gradient-based optimizers in our experiments.
To establish a baseline for comparison, we first examined the optimal learning rate parameter for the GD optimization algorithm.}
\begin{figure}
     \centering
     \includegraphics[width=0.8\textwidth]{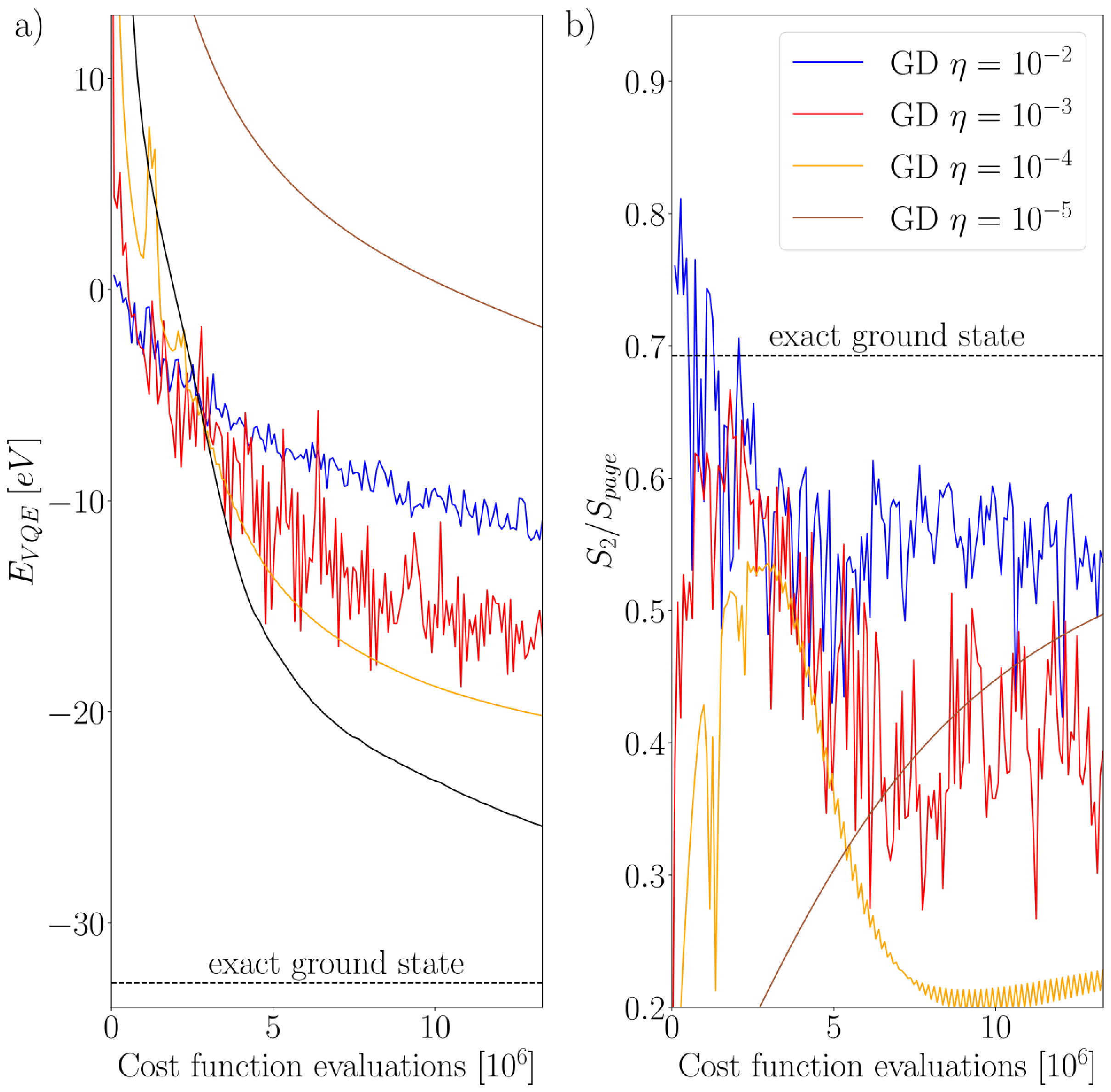} 
     \caption{\pr{Numerical experiments performed with the gradient descent algorithm with various learning rate parameters labeled by $\eta$. In each iterations all of the free parameters are updated.}}
     \label{fig:16qubits_eta}
 \end{figure}
\pr{In Fig.\ref{fig:16qubits_eta}, we visualize the evolution of the VQE energy as a function of cost function evaluation count. We observed pronounced oscillations in the cost function for larger learning rates, specifically for $\eta=10^{-2}$ and $\eta=10^{-3}$ (blue and red lines, respectively). Interestingly, the curve for $\eta=10^{-3}$ exhibits the larger oscillations. This phenomenon can be attributed to the optimization landscape's narrow valleys, which are surrounded by flat territories, consistently with the \prr{BP} theory (i.e. the dominant part of the landscape is expected to be flat, with narrow valleys that are hard to find). A larger learning rate is more likely to overstep these valleys, resulting in oscillations with smaller amplitudes. Conversely, a lower learning rate increases the likelihood of entering these valleys, but may also lead to larger steps during parameter updates (due to the increased gradient inside the valley), causing the algorithm to leave the valley. This reasoning explains the large oscillations in the red line. As the learning rate is further decreased, the oscillations are smoothed out, but the optimization performance is also reduced due to smaller updates in the parameter space (orange and brown lines). Our experiments suggest that learning rates between $\sim10^{-4}$ and $\sim10^{-3}$ are the most effective choice.
}
\begin{figure}
     \centering
     \includegraphics[width=0.8\textwidth]{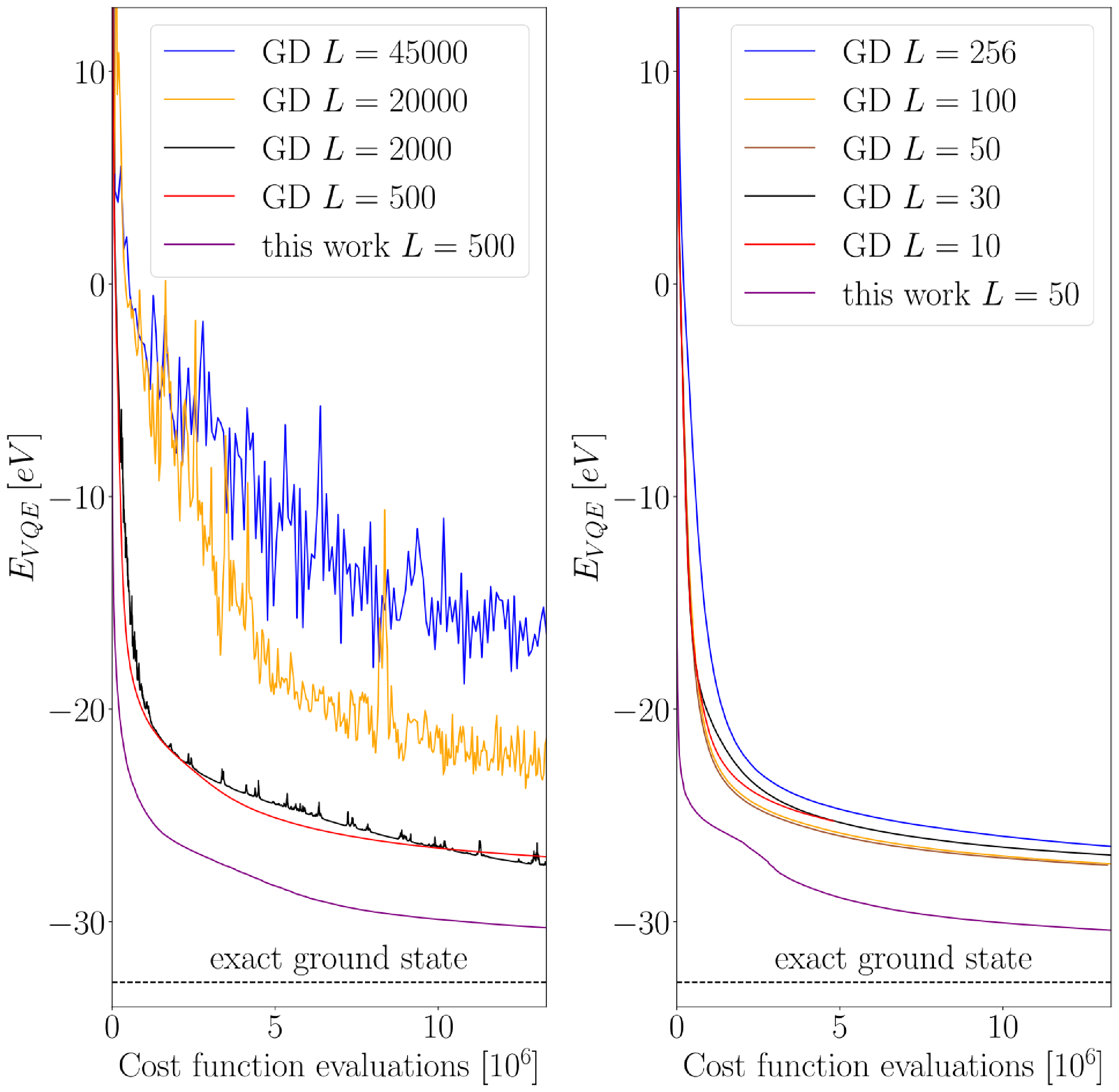} 
     \caption{The efficiency of gradient descent (GD) based VQE optimization for various batch sizes $L$. The batch size refers to the number of parameters taken from the set of the free parameters to be updated during a single GD iteration. In the experiments we applied a learning rate of $\eta=0.001$. 
     The algorithm reported in this work is indicated by the last entry in the legends.}
     \label{fig:16qubits_batch}
 \end{figure}
\pr{Interestingly, the oscillations observed during optimization also disappear when updating only a smaller subset of the parameters in each optimization iteration. 
This means that, similarly to \prr{the} gradient-free multi-parameter search direction approach developed in this work, we randomly select a finite subset of the parameter space in each iteration, evaluate the associated gradient components of the cost function, and apply the parameter update rule on the chosen subset. To illustrate this tendency, Fig.\ref{fig:16qubits_batch} shows the evolution of the VQE energy for various batch sizes $L$. 
(The batch size refers to the number of the randomly chosen components in the  parameters vector.)
The circuit ansatz with $500$ layers has $45000$ free parameters. As we decrease the batch size from $45000$ to $500$, the oscillations in the cost function gradually disappear. While the high dimensionality of the optimization landscape makes it challenging to develop an exact explanation, we can offer a qualitative reasoning for this observation. 
By selecting a small subset of parameters to vary in each iteration, the circuit's expressiveness gets narrowed down, effectively reducing the risk of getting onto a BP when initiated from zero parameters \cite{Holmes2016}.
By optimizing a low expressive circuit in each iteration of the training process one can effectively avoid BPs during the optimization. 
However, this procedure does not fully solve the BP issue, as optimization over a selected subspace of the parameter space can only reach the minimum corresponding to that particular subspace. 
If already close to a local minimum (or on a BP), the improvement in the cost function will be small. 
These two competing aspects predestinates a batch size for which the optimization is the most efficient.
Our numerical results in Fig.\ref{fig:16qubits_batch} suggest that a batch size around $50$ is the most efficient choice for gradient descent optimization.

Now, we'd like to discuss the impact of parameter initialization on the optimization process. 
As discussed previously, when starting from a randomly initialized parameter set, the improvement from optimizing a randomly chosen parameter subspace is in general small. 
This is expected, as the starting point is likely to be located on a BP. 
In contrast, when starting the optimization with zero initialization (i.e., setting the single qubit rotation gates to identity operations), the VQE energy undergoes a much larger improvement during the first stage of the optimization. 
This is what we have exploited in our numerical results shown in Fig.\ref{fig:16qubits_batch} as well. 
To avoid zero gradient components at the beginning of the optimization, we set the initial parameters to randomly selected, but small values close to zero. Our gradient-free optimization strategy can be started from exact zero initial parameters, allowing for even larger initial decrements in the VQE energy, as plotted by the purple lines of Fig.\ref{fig:16qubits_batch}. 
After a pronounced initial decrement in the cost function, the degree of further improvements in the VQE energy becomes modest, indicating the impact of BPs surrounding the global minimum of the optimization problem. 
The second-Rényi entropy of the exact ground state is relatively high [see Fig.~\ref{fig:16qubits}.b)], suggesting the presence of a BP.}


Our numerical results highlight two main conclusions:
(i) Firstly, the probabilistic algorithm developed in this work, making use of the global properties of the optimization landscape, is a promising candidate capable of solving variational quantum problems at scale, even beyond the capabilities of classical quantum computer simulators.  
(ii) Secondly, when using a general circuit ansatz to perform VQE optimization (like the circuit structure used in our experiments), highly accurate results require an exponentially scaling number of layers in the circuit ansatz. We acknowledge that the circuit ansatz used in our calculations may not be the most efficient one to model the ground states of the studied models. Ansatz structures accounting for the symmetry of the addressed physical model might significantly reduce the circuit depth for practical applications \cite{ogorman2019generalized,Anselmetti_2021,Kottmann2023molecularquantum}. Our aim in this work was to examine the numerical capabilities of the new solver strategy. The non-efficient representation of the circuit ansatz did not impose limitations on us in achieving this goal. On the contrary, we could execute quite deep circuits and demonstrate that our strategy can be successfully applied even under extreme circumstances.

\paragraph*{Quantum compilation of deep circuits:}

A commonly employed approach in the compilation and optimization of large quantum circuits involves partitioning the circuit into smaller blocks that can be individually optimized or re-synthesized. This allows for the treatment of wide quantum circuits, as demonstrated in the work of Ref.\cite{kukliansky2023qfactor}, where circuits ranging from $4$ to $400$ qubits were addressed. Previous approaches following this partitioning strategy \cite{9743148} were shown to scale to similar circuit sizes but were practically limited to using only $3$-qubit sized partitions due to inefficient numerical optimization. In the experiments of Ref.\cite{kukliansky2023qfactor}, this limitation was overcome, extending the partition size to $6$ qubits. The choice of the number of qubits per partition is crucial, as more qubits are likely to capture additional domain-level physical interactions, leading to improved optimization opportunities.

On the contrary, for partitions containing more qubits, the optimization problem becomes exceptionally challenging for traditional gradient-based solvers due to the dense presence of local minima in the landscape \cite{squander2,RAKYTA2024112756}. 
Here, we demonstrate that the evolutionary optimization algorithm described in this study enables the application of the adaptive gate synthesis approach of Ref. ~\cite{squander2} to problems that were found to be intractable with conventional optimization engines, such as the BFGS method \cite{squander2}.
In order to measure the distance between the initial unitary $U$ of the quantum program and the synthesized unitary $V$ (obtained from the synthesized circuit) we used the frobenius norm given by
\begin{equation}
    f(U,V) = \frac{1}{2}\left\|V-U \right\|_F^2 = d - \textrm{Re}\left[{\textrm{Tr}}(U^{\dagger}V)\right], \label{eq:frobenius}
\end{equation}
with $d$ being the dimension of the Hilbert-space.
For further details of the decomposing algorithm please refer to either Ref.~\cite{squander2} or to Ref.\cite{RAKYTA2024112756}.
\begin{table*}[ht!]
\centering
\begin{tabular}{|| c | c |c | c | c |  c | c | c ||} 
 \hline
  Circuit name  & qubits & Initial &  QISKIT & \multicolumn{2}{c|}{Ref \cite{squander2,RAKYTA2024112756}} & \multicolumn{2}{c||}{SQUANDER\cite{SQUANDER_github}} \\
       & & $CNOT$ & $CNOT$ & $CNOT$  & $\overline{T}\,[s]$ & $CNOT$ & $\overline{T}\,[s]$ \\ [0.5ex] 
 \hline\hline
 4gt5\_76 & $5$ & $46$ & $529$ &  $24$ & $1711$ & $22$ & $79$  \\
  4gt10-v1\_81 & $5$ &  $66$ & $372$ & $39$ & $65737$ & $37$ & $2245$  \\
  one-two-three-v1\_99 & $5$ &  $59$ & $302$ & $45$ & $80390$  & $35$ & $4693$  \\
  one-two-three-v0\_98 & $5$ &  $65$ & $213$ & $61$ & $175994$  & $42$ & $20313$  \\
one-two-three-v2\_100 & $5$ &  $32$ & $502$ & $37$ & $5141$ & $27$ & $2687$  \\  
  4mod7-v1\_96 & $5$ &  $72$ & $150$ & $33$ & $10255$ & $26$ & $2951$  \\
  aj\_e11\_165 & $5$ &  $69$ & $337$ & $36$ & $15585$ & $23$ & $2823$  \\
  alu-v2\_32 & $5$ &  $72$ & $469$ & $41$ & $33820$ & $31$ & $1639$  \\
  alu-v4\_36 & $5$ &  $51$ & $193$ & $40$ & $11090$ & $28$ & $2106$  \\
4gt5\_77 & $5$ &  $58$ &  $338$ & $19$ & $2855$ & $17$ & $483$  \\
4gt12-v0\_86 & $6$ &  $116$ &  $1837$ & - & - & $40$ & $8174$  \\
4gt12-v0\_87 & $6$ &  $112$ &  $1837$ & $47$ & - & $36$ & $8598$  \\
4gt12-v0\_88 & $6$ &  $86$ &  $1837$ & $44$ & - & $36$ & $6577$  \\
\hline
\end{tabular}
\caption{$CNOT$ gate count comparison of deep $4-6$-qubit circuits obtained by decomposing and optimizing unitaries taken from the online database \cite{ibm_mapping}. The individual columns label the name of the circuit, the number of qubits in the circuit, the initial $CNOT$ gate count of the circuit, the $CNOT$ gate count of the quantum circuit synthesized by QISKIT transpile function. The last four columns show the $CNOT$ count and average execution times obtained from Refs.~\cite{squander2} and \cite{RAKYTA2024112756}, and the same kind of results obtained with the new evolutionary algorithm with the SQUANDER package. The final error $f$ of the approximation calculated via Eq.~(\ref{eq:frobenius}) was less than $10^{-8}$ in each decomposition corresponding.
We performed the benchmark calculations on a computing server equipped with $32$-Core AMD EPYC 7542 Processor (providing $64$ threads with multi-threading) and with $128$GB of memory.}
\label{table:IBM_CNOT_gates_opt}
\end{table*}
Our numerical experiences revealed that quantum gate synthesis provides a perfect opportunity to explore problems where the presence of local minima in the optimization landscape causes traditional optimizers to fail, even at smaller scales. In our experiments, we observed that employing the developed evolutionary algorithm with multiple agents is indeed necessary to synthesize deep circuits while achieving a significant reduction in gate count. We executed $64$ concurrent agents and synchronizing their evolution after $200-2000$ optimization steps. To reduce the computational footprint, we limited the randomly chosen set of parameters to contain only a single parameter. Nevertheless, the developed evolutionary algorithm demonstrated exceptional efficiency in solving the addressed problems. For optimal performance, we combined the evolutionary algorithm with the BFGS strategy, switching to the latter when the cost function was reduced down to $10^{-3}$ by the evolutionary algorithm. The results of our numerical experiments are presented in Table.~\ref{table:IBM_CNOT_gates_opt}. Comparing these results with previous findings, we can draw two conclusions. Firstly, the average execution time was reduced by an order of magnitude in all addressed use cases where comparisons could be made. Secondly, the evolutionary algorithm exhibited remarkable stability and efficiency in solving the optimization problem with a precision of $10^{-8}$.

\paragraph*{Complexity considerations:}

It is worth to study the computational complexity of the developed evolutionary algorithm described in this study.
There are two key components in relation with our optimization technique characterizing it's cost. 
These are the utilization of the agents undergoing evolutionary selections and the size of the parameter subset over which the cost function gets optimized.
Both of these dimensions, though, fall back to the count of sampling from  the single parameter cross sections of the cost function.
Since the terminating point of the line search can be reused in the next iteration, only $2$ cost function evaluations are needed to reconstruct the parameter dependence of the cost function, if Eq.~(\ref{eq:3point}) is used to characterize the optimization landscape.
In the more general case of Eq.~(\ref{eq:5point}) $4$ cost function evaluations are required to determine the minimum with respect to a single parameter.
Therefore, $2LM$ (or $4LM$) cost function evaluations are required to determine the search direction, and additional constant number of evaluations to perform the line search.
These numbers compare close to the case of regular gradient descent method.
In particular, on a quantum computer, the general rule of parameter shift \cite{Wierichs2022generalparameter} can be used to evaluate a gradient component of the cost function with $2$ evaluations of the cost function.
By carefully planning the number of agents and the number of optimized parameter components per iteration, the overall complexity of our optimization algorithm can be set not to exceed the complexity of conventional gradient-based methods.

\section*{Discussion}  \label{sec:conclusion}

In this manuscript, we have presented a novel optimization method for training quantum circuits. A prevalent challenge in variational quantum algorithms is the hindering impact of BPs on circuit training. The vanishing gradient problem associated with BPs significantly limits the applicability of near-term quantum algorithms. Clearly speaking, the application of a given variational algorithm becomes questionable when the training process fails at large scaled problem instances.
Therefore, developing optimization strategies resistant to BPs is highly desirable.
We demonstrated that the long-range properties of the optimization landscape can be leveraged to determine a search direction and optimization bounds along this direction. This method introduces a novel \pr{batched} line-search approach for training quantum circuits, autonomously maintaining low values of the second-Rényi entropy during optimization, \pr{provided the optimization starts with zero qubit rotation initialization.} Given the inherent connection between entropy and BPs, this property of the developed algorithm implies a natural tendency to avoid BPs during training without requiring external control mechanisms. \pr{This is achieved by effectively reducing the expressibility (or variability) of the circuit ansatz via updating only a smaller subset of the optimization parameters in each iteration. 
Compared to traditional gradient based optimization approaches -- like the plain gradient descent or the ADAM algorithm -- our numerical results yielded an increased efficiency of our optimization approach during the first periods of the optimization process, where our method showcased significantly larger improvements on the cost function. 
This is where our optimization method delivers advantage, as apparently bigger improvement can be achieved within fewer cost function evaluations.
Powell's gradient free optimization method is the one we should also mention here, as our approach shares some similarities with this method. 
Therefore, the numerical experiments performed with Powell's optimization method got quite close to our results. 
Our approach is capable of outperforming Powell's method by two means. 
Firstly, our method adopts the properties of the optimization landscape in making the updates of the parameters.
Secondly, in contrast to Powell's method, our approach can be parallelized, significantly shortening the training time.}

We applied the optimization strategy to VQE simulations on randomly generated Heisenberg XXX and Sachdev–Ye–Kitaev (SYC) models, all of which were previously identified to exhibit barren plateaus (BPs) during the VQE process. We increased the depth of the training circuits to $p=1000$ entangling layers with $16$ qubits \prr{and to $p=500$ entangling layers with $21$ qubits}. The developed optimization strategy, utilizing the gradient-free search direction method, consistently decreased the VQE energy over thousands of iterations without getting trapped by BPs.
We demonstrated these outstanding solver capabilities by training quantum circuits containing up to $15000$ entangling gates.

In this study we also addressed another common issue of optimization approaches:
the local minima scattered over the optimization landscape causes early termination of the optimization and prevents from solving the addressed problem. 
This issue is particularly critical for quantum gate synthesis, where only results close to the global minimum are acceptable. Our numerical experiments demonstrated a significant increase in efficiency to avoid local minima and converge to the global minimum when employing evolutionary selection of agents. In most of the studied circuits, we achieved a substantial reduction in gate count and decreased the computational time by an order of magnitude. For these experiments, we utilized single-parameter optimization to keep computational requirements low.

We believe that the optimal training performance for large-scale variational quantum calculations can be achieved through the combination of these practices. 
However, due to the exponential scaling of quantum circuit simulation with the number of qubits, the limitations of classical computers have prevented us from conducting exhaustive research by applying the gradient-free multi-parameter search direction strategy and evolutionary selection in combination.

Combining the described probabilistic optimization strategy with machine learning practices has further potential to further improve its performance. Replacing the probability-based selection with machine learning inference could significantly reduce the number of redundant evaluations of the cost function during circuit training. We leave the exploration of this possibility for future research projects.

\prr{This work focuses on the developing and testing optimization strategies in an ideal, noise-free environment to understand their theoretical behavior before tackling the effects of hardware imperfections during the circuit training.
We plan an extensive study on optimal circuit shot counts, which we expect to vary throughout the training and depend on other hyperparameters such as the batch size.
Additional effects like qubit relaxation or amplitude/phase damping may also impact performance, and we see their investigation as a substantial project in its own right.
Such future work will build on the present theoretical foundation to prepare the method for noisy, real-world quantum hardware.
}

\section*{Acknowledgments}

\textbf{Founding:} This research was supported by the Ministry of Culture and Innovation and the National Research, Development and Innovation Office within the Quantum Information National Laboratory of Hungary (Grant No. 2022-2.1.1-NL-2022-00004), by the ÚNKP-23-5 New National Excellence Program of the Ministry for Culture and Innovation from the source of the National Research, Development and Innovation Fund, and by the Hungarian Scientific Research Fund (OTKA) Grants No. K134437 and FK135220.
We also acknowledge the support from the QuantERA II project HQCC-101017733 \prr{and from the Horizon Europe programme HORIZON-CL4-2022-QUANTUM-01-SGA via the project 101113946 OpenSuperQPlus100.}
RP. acknowledge support from the Hungarian Academy of Sciences through the Bolyai J\'anos Stipendium (BO/00571/22/11) as well.
We acknowledge the computational resources provided by the Wigner Scientific Computational Laboratory (WSCLAB) (the former Wigner GPU Laboratory).

\bibliographystyle{quantum}
\bibliography{references}

\onecolumn\newpage
\appendix

\section{Technical details on the derivation of Eqs.~(\ref{eq:3point}) and (\ref{eq:5point})}

\pr{The derivation of Eqs.~(\ref{eq:3point}) and (\ref{eq:5point}) relies on the analytical properties of the quantum gates incorporated in the trained circuits.
In general, each of the parametric gates can be expressed via trigonometric functions of its parameter, namely:
\begin{equation}
 U = \cos(p) U_1 + \sin(p) U_2 + U_3\;, \label{eq:repr}
\end{equation}
where constant matrices $U_i$ are of the same size as the matrix of the gate operation labeled by $U$.
For single qubit rotations (including RX, RY, RZ) the matrix $C$ is zero, while for constant gates (like the $CNOT$ gate) matrices $A$ and $B$ are zeros.
By applying this gate on an initial state $|0\rangle$, we arrive at the transformed state
\begin{equation}
    |\Psi\rangle  = U_4 \times \bigg(\cos(p) U_1 + \sin(p) U_2 + U_3\bigg) \times U_5 |0\rangle\;,
\end{equation}
where $U_i$'s are constant matrices with respect to the chosen parameter $p$.
This expression can be further simplified to 
\begin{equation}
    |\Psi\rangle  = \cos(p)|a\rangle + \sin(p)|b\rangle + |c\rangle
\end{equation}
with respect to the parameter $p$, where $|a\rangle$, $|b\rangle$ and $|c\rangle$ are suitable states from the Hilbert space. 
By evaluating the expectation value of the Hamiltonian we get:
\begin{eqnarray}
 \langle\Psi |\hat{H}| \Psi \rangle &=& \cos(p)^2 H_{aa} + \sin(p)^2 H_{bb} + H_{cc} + \\
 &+&\cos(p)\sin(p)\left(H_{ab}+H_{ba}\right) + \\
 &+&\cos(p)(H_{ac}+H_{ca}) + \sin(p) (H_{bc}+H_{cb} )\;, 
\end{eqnarray}
where matrix elements $H_{ij}$ ($i,j\in\{a,b,c\}$) are defined as $H_{ij} = \langle i |\hat{H}| j \rangle$.
After utilizing trigonometric identities $2\cos(p)^2 = \cos(2p)+1$ and $2\sin(p)\cos(p)=\sin(2p)$, the equation simplifies to:
\begin{eqnarray}
 \langle\Psi |\hat{H}| \Psi \rangle &=& \cos(2p)\frac{H_{aa}-H_{bb}}{2} + \sin(2p)\frac{H_{ab}+H_{ba}}{2} \\
 &+& \cos(p) H_{ac} + \sin(p) H_{bc} + \frac{H_{ba}+H_{bb}}{2} + H_{cc}\;.
\end{eqnarray}
Since we can always transform the expression $\cos(p)c_1 + \sin(p)c_2$ into a form of $A\cos(p+\varphi)$ with $A = \sqrt{(c_1)^2+(c_2)^2}$ and $\sin(\varphi) = c_1/A$, we can transform the equation above into the form of
\begin{equation}
  E_{VQE} = \kappa\cdot\sin(2p_i+\xi) + \gamma\cdot\sin(p_i+\phi) + C \;,
\end{equation}
were the parameters $\kappa$, $\xi$, $\gamma$, $\phi$ and $C$ are determined by the Hamiltonian elements $H_{ij}$ ($i,j\in\{a,b,c\}$).
When the (\ref{eq:repr}) representation of the selected gate contains no constant $U_3$, then we arrive at the expression:
\begin{equation}
 E_{VQE} = \kappa\cdot\sin(2p_i+\xi) + C \;.
\end{equation}
These equations match Eqs.~(\ref{eq:3point}) and (\ref{eq:5point}) in the main text.}

\end{document}